\documentclass{aa}  

\usepackage{graphicx}
\usepackage{txfonts}
\usepackage{lipsum}
\usepackage{subcaption}         % necessary for continued figures, example in section 3
\usepackage{lscape}             % to rotate a single page table, example in appendix.
\usepackage{placeins}           % useful with \FloatBarrier, to keep 
\usepackage{natbib}
\usepackage{multirow}
\AddToHook{begindocument/before}{\RequirePackage{hyperref}}
\usepackage[colorlinks=true, linkcolor=blue, citecolor=blue, filecolor=blue, urlcolor=blue]{hyperref}

\begin{document}

\title{Coordinated Coronagraphic Observations from Proba-3 and Aditya-L1: Investigating CME Energetics in the Inner Corona}

\author{Anjali Agarwal\inst{1,2}\corrauth{anjaliagarwal1024@gmail.com}       
\and Wageesh Mishra\inst{1,2}\corrauth{m.wageesh30@gmail.com}
\and R. Ramesh\inst{1}\email{ramesh@iiap.res.in}
\and V. Muthu Priyal\inst{1}\email{muthu.priyal@iiap.res.in}
\and A.N. Zhukov\inst{3}\email{Andrei.Zhukov@sidc.be}
\and Jagdev Singh\inst{1}\email{jsingh@iiap.res.in}
\and Nat Gopalswamy\inst{4}\email{natchimuthuk.gopalswamy-1@nasa.gov}
\and Soumyaranjan Khuntia\inst{1}\email{khuntias133@gmail.com}
\and K. Sasikumar Raja\inst{1}\email{sasikumar.raja@iiap.res.in}
\and B. Bourgoignie\inst{3}\email{bram.bourgoignie@oma.be}
\and L. Dolla\inst{3}\email{laurent.dolla@oma.be}
\and S. Shestov\inst{3,5}\email{s.shestov@oma.be}
\and S. Fineschi\inst{6}\email{silvano.fineschi@inaf.it}
\and S. Gunar\inst{7}\email{stanislav.gunar@asu.cas.cz}
\and P. Lamy\inst{8}\email{philippe.lamy@latmos.ipsl.fr}
\and M. Mierla\inst{3,9}\email{marilena.mierla@oma.be}
\and H. Peter\inst{10}\email{peter@mps.mpg.de}
\and P. Rudawy\inst{11}\email{rudawy@astro.uni.wroc.pl}
\and K. Tsinganos\inst{12}\email{tsingan@phys.uoa.gr}}

\institute{Indian Institute of Astrophysics, II Block, Koramangala, Bengaluru 560034, India
\and Pondicherry University, R.V. Nagar, Kalapet 605014, Puducherry, India
\and Solar–Terrestrial Centre of Excellence — SIDC, Royal Observatory of Belgium, 1180 Brussels, Belgium
\and Heliophysics Science Division, NASA Goddard Space Flight Center, Greenbelt, MD 20771, USA
\and Centre Spatial de Liège, Université de Liège, Av. du Pré-Aily B29, 4031 Angleur, Belgium
\and National Institute for Astrophysics, Astrophysical Observatory of Torino, Pino Torinese, Torino, Italy
\and Astronomical Institute of the Czech Academy of Sciences, 251 65 Ondřejov, Czech Republic
\and Laboratoire Atmosphères et Observations Spatiales, 11 Boulevard d’Alembert, 78280 Guyancourt, France
\and Institute of Geodynamics of the Romanian Academy, 020032, Bucharest-37, Romania
\and Max Planck Institute for Solar System Research, Justus-von-Liebig-Weg 3, 37077, Göttingen, Germany
\and Astronomical Institute, University of Wrocław, Kopernika 11, 51-622 Wrocław, Poland
\and University of Athens, Panepistimiopolis, 157 84 Zografos Athens, Greece}

\date{Received September 30, 20XX}

\titlerunning{Energetics of CMEs from Proba-3 and Aditya-L1}
\authorrunning{Agarwal et al.}
\abstract
{Constraining the plasma properties and energetics of coronal mass ejections (CMEs) in the inner corona is essential for understanding their early evolution, yet remains challenging because of limited observations.}
{We investigate the mass and density evolution, and energy partitioning of two CMEs observed on 2025 September 14 and 16, and assess the potential of coordinated Proba-3/ASPIICS and Aditya-L1/VELC observations to constrain CME energetics in the low corona.} 
{We present the first coordinated observations of CMEs obtained simultaneously by Proba-3/ASPIICS and Aditya-L1/VELC. Using ASPIICS white-light observations, we estimate the CME mass, volume, number density, and the evolution of kinetic, thermal, and magnetic energies. Magnetic energies are estimated from observed CME properties using observationally constrained, physically motivated assumptions. Simultaneous VELC Fe XIV 5303 $\AA$ observations provide independent estimates of the emission measure, electron number density, thermal energy, and CME lateral extent.}
{The electron number density evolves as $n_e\propto r^{-3.3}$ for the 14 September CME and as $n_e\propto r^{-6}$ for the 16 September CME. The two events exhibit markedly different energy partitioning in the low corona. For the 14 September CME, the kinetic and magnetic energies are comparable, while the thermal energy remains nearly two orders of magnitude smaller, indicating limited plasma heating. In contrast, the 16 September CME exhibits a substantial thermal-energy enhancement, with thermal energy eventually becoming comparable to kinetic energy. For both CMEs, the estimated magnetic energy remains comparable to or exceeds the kinetic energy over the observed height.}
{Our results demonstrate the scientific potential of synergetic ASPIICS and VELC observations for constraining CME mass, density, and energetics in the inner corona, providing new observational constraints on the early evolution of CMEs.}


\maketitle

% \linenumbers
% \modulolinenumbers[1]

%%%%%%%%%%%%%%%%%%%%%%%%%%%%%%%%%%%%%%%%%%%%%%%%%%%%%%%%%%%%%%
\section{Introduction}
Coronal mass ejections (CMEs) are large-scale expulsions of magnetized plasma from the Sun into the heliosphere, and are the primary drivers of space weather effects \citep{Schwenn2006,Webb2012,Schrijver2015,Temmer2024}. CMEs and their substructures (shock, sheath, and flux rope) exhibit a wide range of physical properties, including morphology, angular width, dimensions, mass, energy, and propagation speed \citep{Forbes2000,Vourlidas2000,Yashiro2004,Vourlidas2010,Mishra2015,Mishra2021a}. These properties of CMEs have been routinely investigated during their evolution by combining white-light coronagraph images with in situ observations to understand their Sun-to-Earth evolution \citep{Howard2009,Mishra2016,Kilpua2017,Mishra2023,Agarwal2025,Agarwal2026,Mishra2026}.

Early dynamics of CMEs can provide a better understanding of their initiation and the sudden release of stored magnetic energy in the corona. The energy release is known to be accompanied by processes leading to a loss of equilibrium in the flux rope system \citep{Forbes2000,Moore2001,Kliem2006,Chen2011,Webb2012}. Although several existing models describe CME initiation and early dynamics in the inner corona ($\sim$1.05-1.5~$R_\odot$) \citep{Forbes2006,Fan2007}, the associated energetics and kinematics--such as mass, velocity, acceleration, kinetic and thermal energy, and the restructuring of magnetic field lines--remain poorly constrained due to the observational limitations of CMEs in the inner corona. However, there has been reasonable focus in the literature for understanding the CMEs kinematics, thermodynamics, and energetics in the middle and higher corona using the benchmark coronagraphic and heliospheric imaging observations aboard SOlar and Heliospheric Observatory (SOHO) and Solar TErrestrial RElations Observatory (STEREO) \citep{Davies2009,Mishra2013,Mishra2015a,Mishra2017,Vourlidas2019,Mishra2021,Agarwal2024,Khuntia2025}.

White-light observations of the solar corona and CMEs at heliocentric distances below $\sim$1.5 $R_{\odot}$ are severely limited. Although total solar eclipses can provide observations of the corona in this region, their rarity and brief duration limit detailed investigations of CMEs \citep{Habbal2010,Boe2020}. Consequently, Extreme Ultraviolet (EUV) observations are often employed as an alternative diagnostic of CMEs in the inner corona. However, EUV emission decreases rapidly with height because its intensity scales approximately with the square of the electron density at a given temperature \citep{Zarro1999,Ciaravella2000,Gopalswamy2000a,Harrison2003,Aschwanden2005}. As a result, continuously tracking CMEs as they undergo rapid dynamical and thermal evolution from the inner to the middle and outer corona remains challenging \citep{Auchere2023}.

Several studies have combined EUV observations of the low corona with white-light coronagraph observations of the middle corona to investigate CME evolution and energetics \citep{Chen2003,Gallagher2003,Bien2011,Emslie2012,Byrne2014,OHara2019}. However, such analyses are subject to uncertainties because EUV and white-light observations probe different physical processes and may not trace the same CME structures \citep{Billings1966,Zarro1999}. Likewise, radio and EUV observations have been employed to investigate CME energetics in the low corona \citep{Ramesh2012,Hannah2013,Sasi2014,Gopalswamy2015a,QZhang2023}, but these diagnostics may sample different plasma populations and generally do not enable direct, continuous measurements of CME mass and energy evolution over an extended height range.

Interestingly, space-based coronagraphs, such as LASCO-C1 on SOHO with fields of view (FoV) of $1.1$–$3~R_\odot$ \citep{Brueckner1995} and COR1 (FoV $1.4$–$4~R_\odot$) on STEREO \citep{Howard2008}, partially filled the observational gap of lower to middle corona. However, their relatively low cadence ($\sim$20–60 minutes for LASCO-C1 and $\sim$5 minutes for COR1), low signal-to-noise ratio, and low dynamic range limit their ability to observe the evolution of fainter and energetic CMEs in the low corona \citep{Srivastava2000,Gallagher2003,Magdalenic2008}. Further, due to the LASCO-C1 limited observing period, CMEs in the inner corona did not receive the required attention. In the present era, the advent of next-generation instruments, such as the Association of Spacecraft for Polarimetric and Imaging Investigation of the Corona of the Sun (ASPIICS) aboard Proba-3 \citep{Zhukov2025,Zhukov2025a,Zhukov2026} and the Visible Emission Line Coronagraph (VELC) aboard Aditya-L1 \citep{Singh2019,Ramesh2024,Ramesh2025,Muthupriyal2025,Muthupriyal2025a}, has opened new opportunities to investigate CME energetics in the low corona using simultaneous white-light coronagraphic and spectroscopic observations.

In particular, ASPIICS observes the solar corona over a height range of $\sim$1.1–3~$R_\odot$, effectively bridging the long-standing observational gap between the low and middle corona in white light. ASPIICS operates in six filters corresponding to three spectral channels: a broad white-light band ($5363-5658~\AA$), a narrow band centered on the He~I D3 line ($5877~\AA$), and a narrow band centered on the Fe~XIV green line ($5304~\AA$). The broad white-light channel includes four filters that provide one total-brightness image and three polarized-brightness images. The Fe~XIV green line traces coronal plasma at temperatures of $\sim$2~MK. With significantly reduced stray light compared to earlier coronagraphs, together with cadences of $\sim$30~s in white light and $\sim$5~min in the spectral-line channel, ASPIICS provides unprecedented temporal and spatial coverage of CME and their substructures (shock, sheath, and flux rope) in the low corona. CME shock and its formation height have traditionally been inferred from radio observations \citep{Gopalswamy2009b,Ramesh2012,Gopalswamy2013}, and this could now be possible for fast CMEs using ASPIICS white-light imaging observations, as occasionally reported in earlier studies \citep{Gopalswamy2010,Gopalswamy2011,Gopalswamy2012a}. It is important to note that ASPIICS imaging observations can be complemented with VELC spectroscopic (Fe~XIV green line $5303~\AA$) observations (over a height range of $\sim$1.05–1.5~$R_\odot$) as they have a common FoV in the low corona.

Combining observations from ASPIICS and VELC provides a unique opportunity to estimate CME energetics. In particular, VELC is the first space-based instrument capable of routinely observing CMEs in the Fe~XIV green line at critical inner-coronal heights \citep{Singh2019,Ramesh2024}. Motivated by these capabilities and by the lack of previous studies combining ASPIICS and VELC observations, we present the first coordinated analysis of ASPIICS and VELC observations for two selected CMEs. Using the ASPIICS wideband and Fe~XIV imaging channels, we investigate the kinematics, mass, and density evolution of CMEs in the inner corona and estimate their kinetic, thermal, and magnetic energies, with the magnetic-energy estimates derived using observationally constrained and physically motivated methods. To further constrain the plasma properties, we derive independent estimates of CME mass and thermal energy from VELC spectroscopic observations and compare them with the corresponding ASPIICS results. The event selection, CME tracking, and kinematic analysis are described in Section~\ref{sec:cme_kinematics}. The mass estimates from ASPIICS and VELC observations are presented in Sections~\ref{sec:cme_parameters} and \ref{sec:velc}, respectively, while the estimation and comparison of CME energetics are discussed in Section~\ref{sec:comparing_energy}. Section~\ref{sec:discussion} discusses the main results and the limitations of the study, and Section~\ref{sec:conclusion} summarizes the main conclusions.

\begin{figure*}
\centering
\includegraphics[scale= 0.211,trim={0cm 0cm 0cm 0cm},clip]{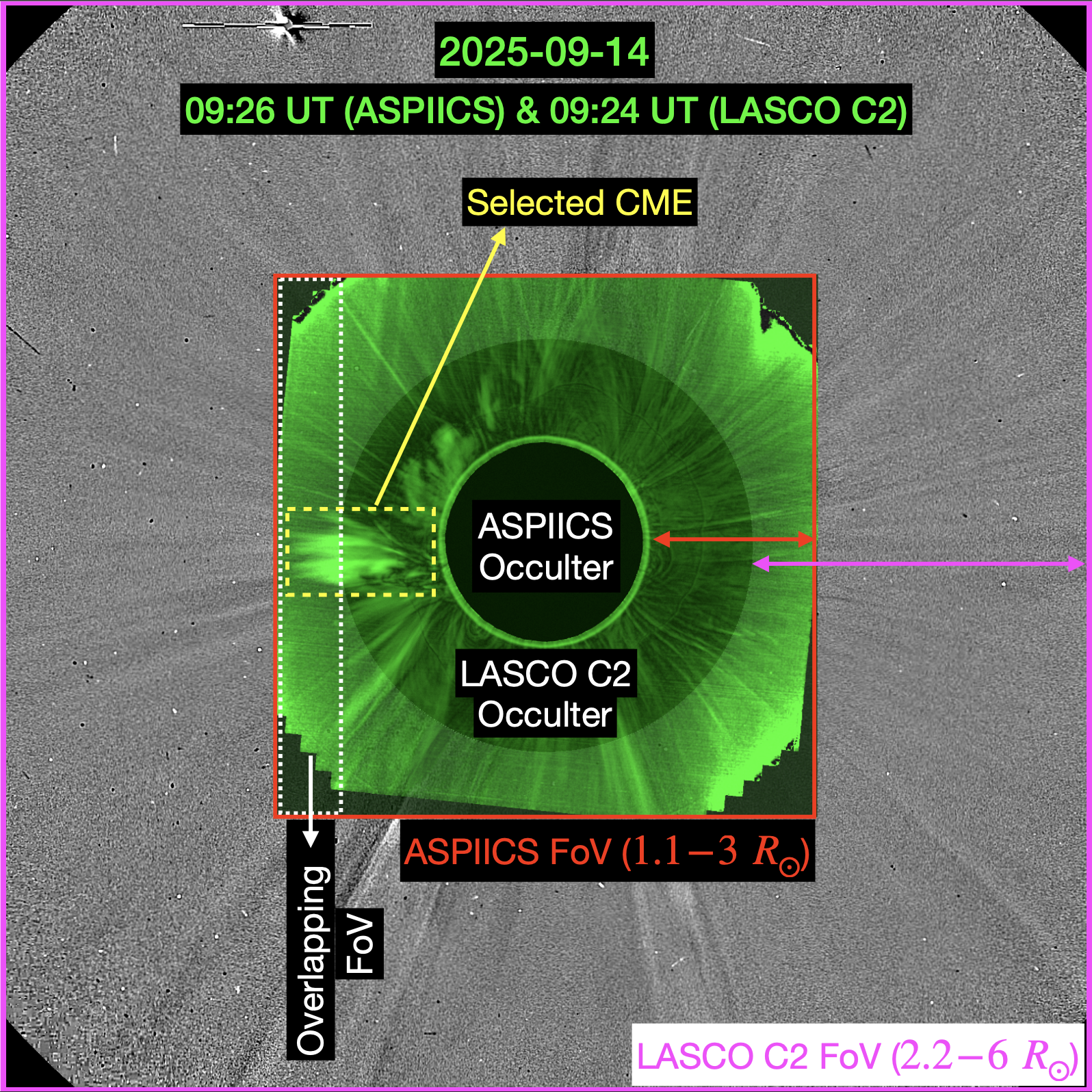}
\includegraphics[scale= 0.211,trim={0cm 0cm 0cm 0cm},clip]{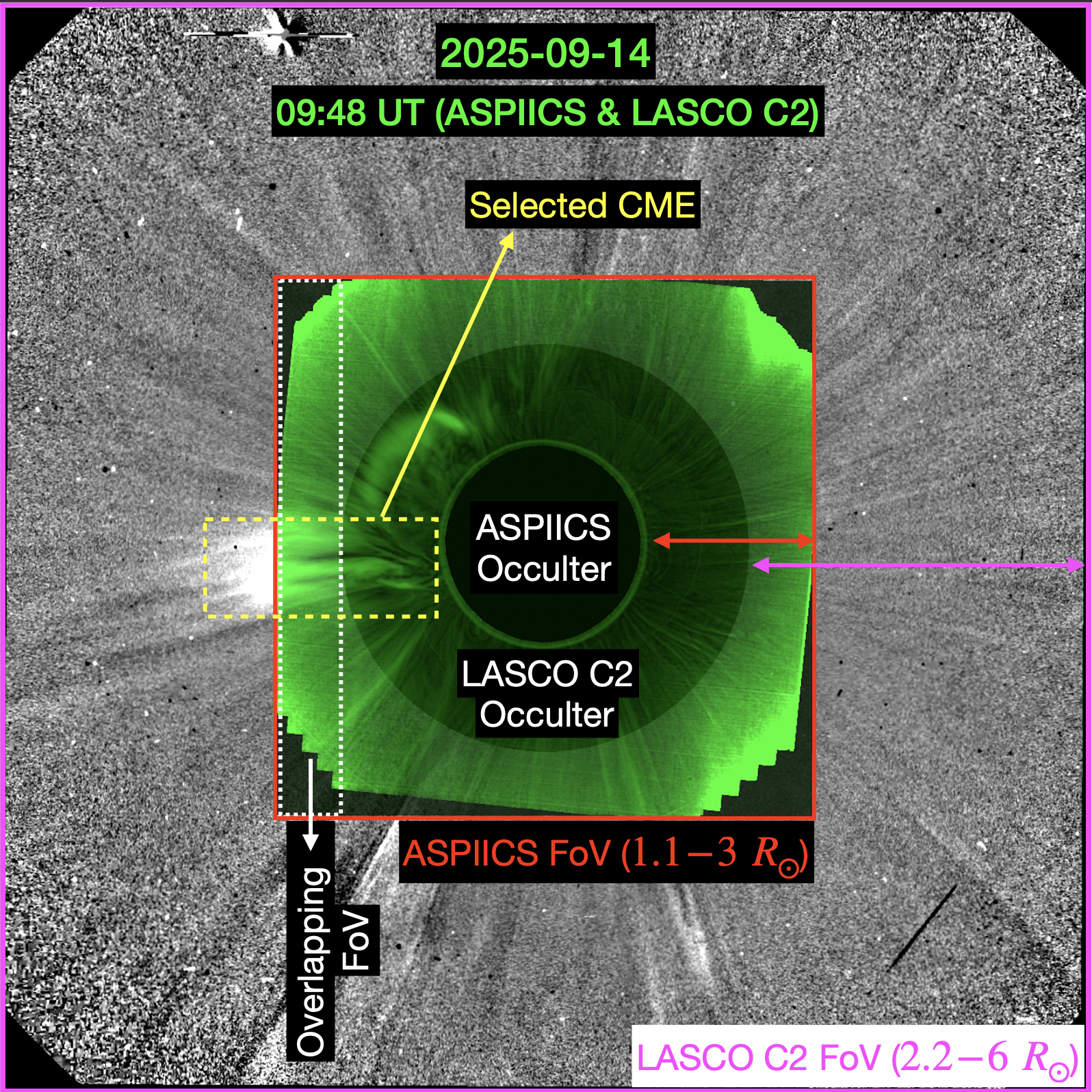}
\includegraphics[scale= 0.211,trim={0cm 0cm 0cm 0cm},clip]{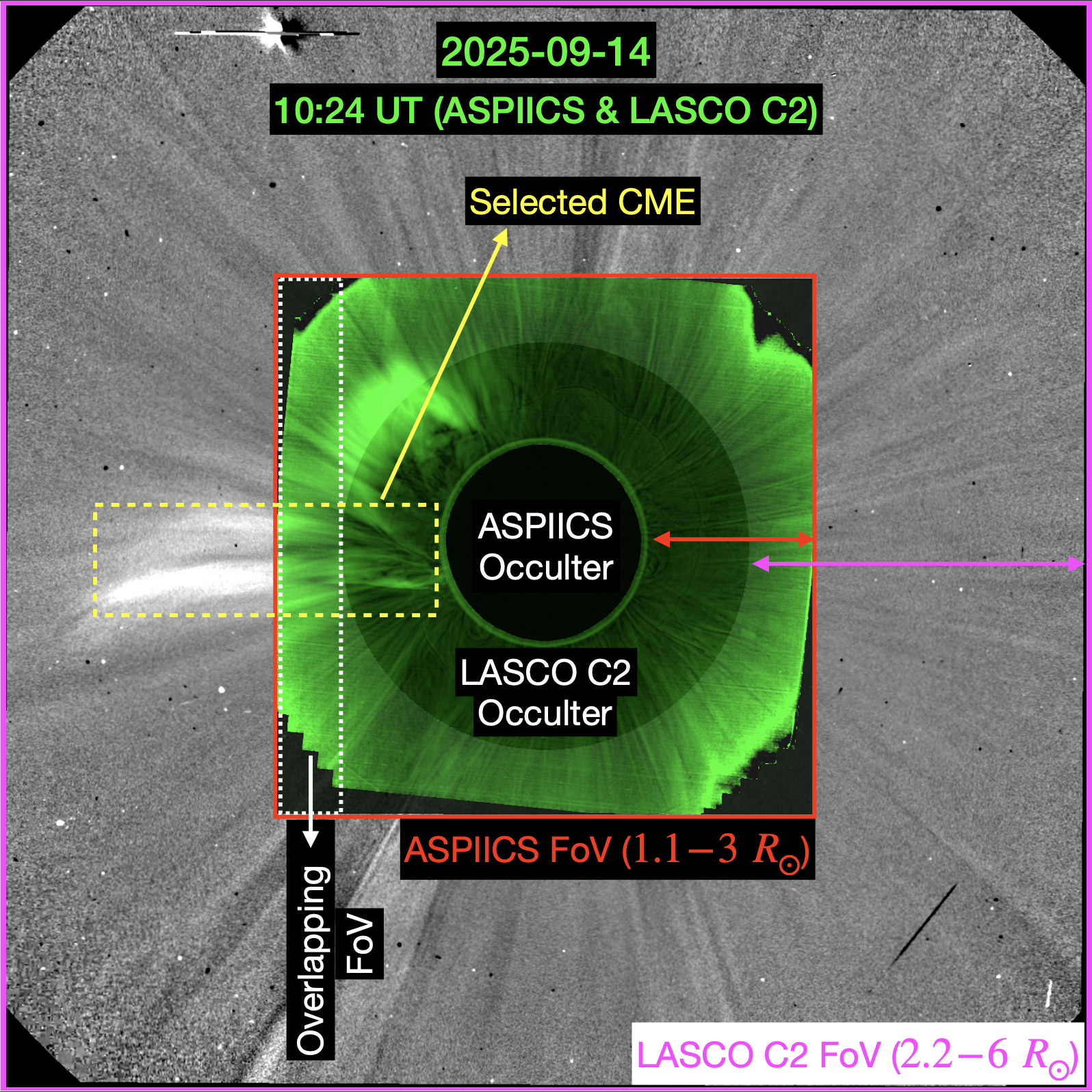}\\[0.3cm]
\includegraphics[scale= 0.211,trim={0cm 0cm 0cm 0cm},clip]{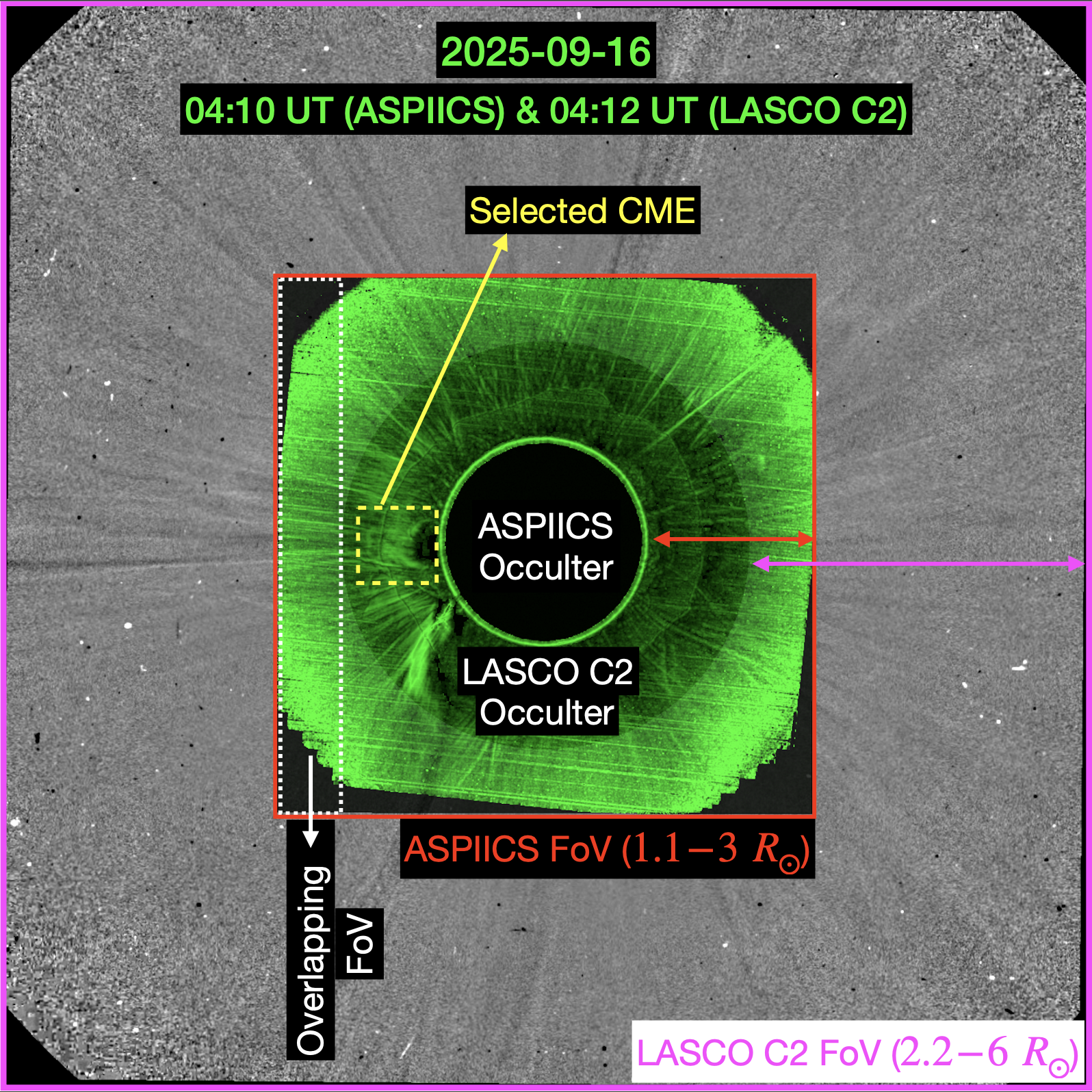}
\includegraphics[scale= 0.211,trim={0cm 0cm 0cm 0cm},clip]{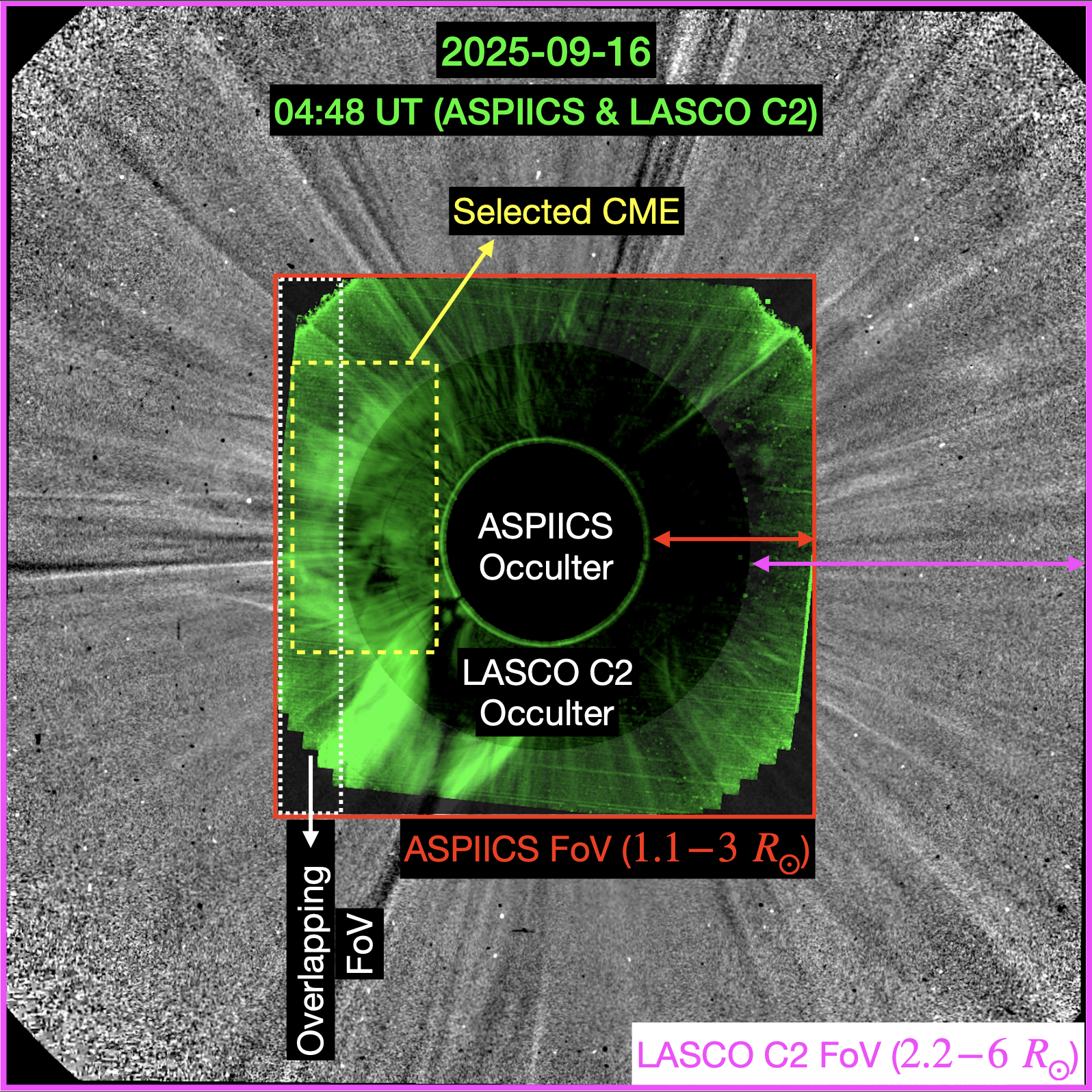}
\includegraphics[scale= 0.211,trim={0cm 0cm 0cm 0cm},clip]{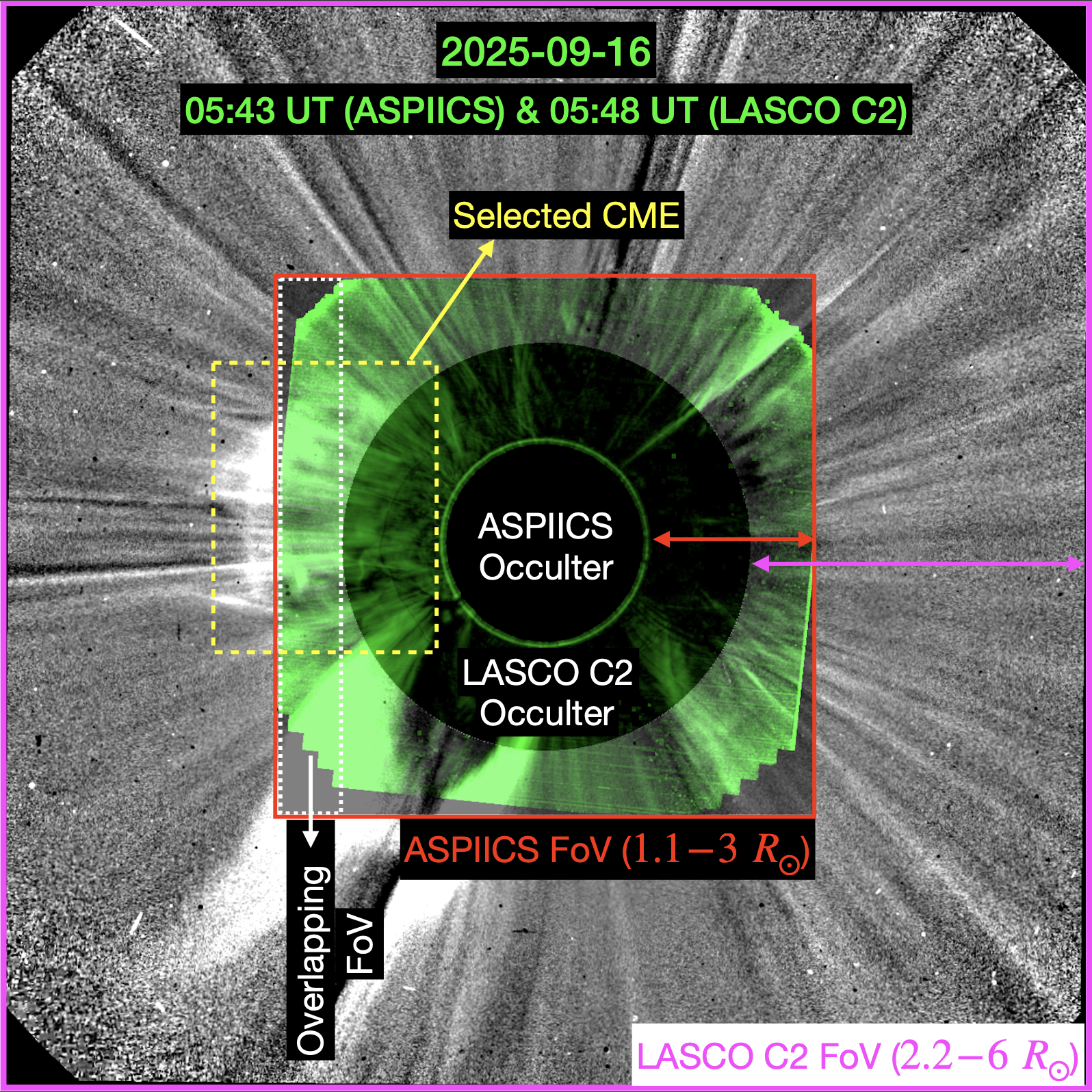}
\caption{Composite of ASPIICS base-difference images (FoV outlined in red) and LASCO-C2 running-difference images (FoV outlined in magenta) for the 14 Sep (top panel) and 16 Sep (bottom panel) CMEs at three time instances. The selected CMEs are marked by yellow dashed boxes, while the overlapping FoV between ASPIICS and LASCO-C2 is indicated by white dotted boxes.}
\label{fig:composite_image}
\end{figure*}

\section{Observations, Properties, and Energetics of Selected CMEs Observed by ASPIICS and VELC Coronagraphs} 
\label{sec:obs}

Coordinated observations from the ASPIICS and VELC coronagraphs provide a unique opportunity to investigate CME mass, number density, and energetics in the inner corona ($\lesssim3~R_{\odot}$), a critical region for the peak acceleration of CMEs \citep{Srivastava2000,Gallagher2003,Zhang2006}. Simultaneous observations from these two coronagraphs, with overlapping fields of view, enable direct cross-validation of the derived CME properties and assessment of methodological uncertainties. For this purpose, we first compiled a list of CME events observed by ASPIICS during the first six months of its nominal mission phase (July--December 2025). We identified more than ten CMEs in the ASPIICS observations. However, owing to the limited overlap in the observing duty cycles of ASPIICS and VELC, only two events, observed on 2025 September 14 and 16, were recorded simultaneously by both instruments.

These CMEs were noted to first appear in LASCO-C2 images at approximately 09:12~UT on 14 Sep and 04:24~UT on 16 Sep 2025, respectively, according to the CDAW CME catalog (\url{https://cdaw.gsfc.nasa.gov/CME_list/}). The early detections of these CMEs in ASPIICS and VELC enable their reliable association with the corresponding evolved structures observed later in the LASCO FoV. This further underscores the importance of ASPIICS observations for probing the inner corona, a height range not fully accessible to LASCO. Figure~\ref{fig:composite_image} shows composite ASPIICS and LASCO-C2 coronagraph images for the 14 Sep (top panel) and 16 Sep (bottom panel) CME events at three representative time instances. The selected CMEs are marked by yellow dashed boxes, while the overlapping field of view between ASPIICS and LASCO-C2 is indicated by white dotted boxes, illustrating the correspondence between the two instruments.

\begin{figure*}
\centering
\includegraphics[scale= 0.215,trim={0cm 0cm 0cm 0cm},clip]{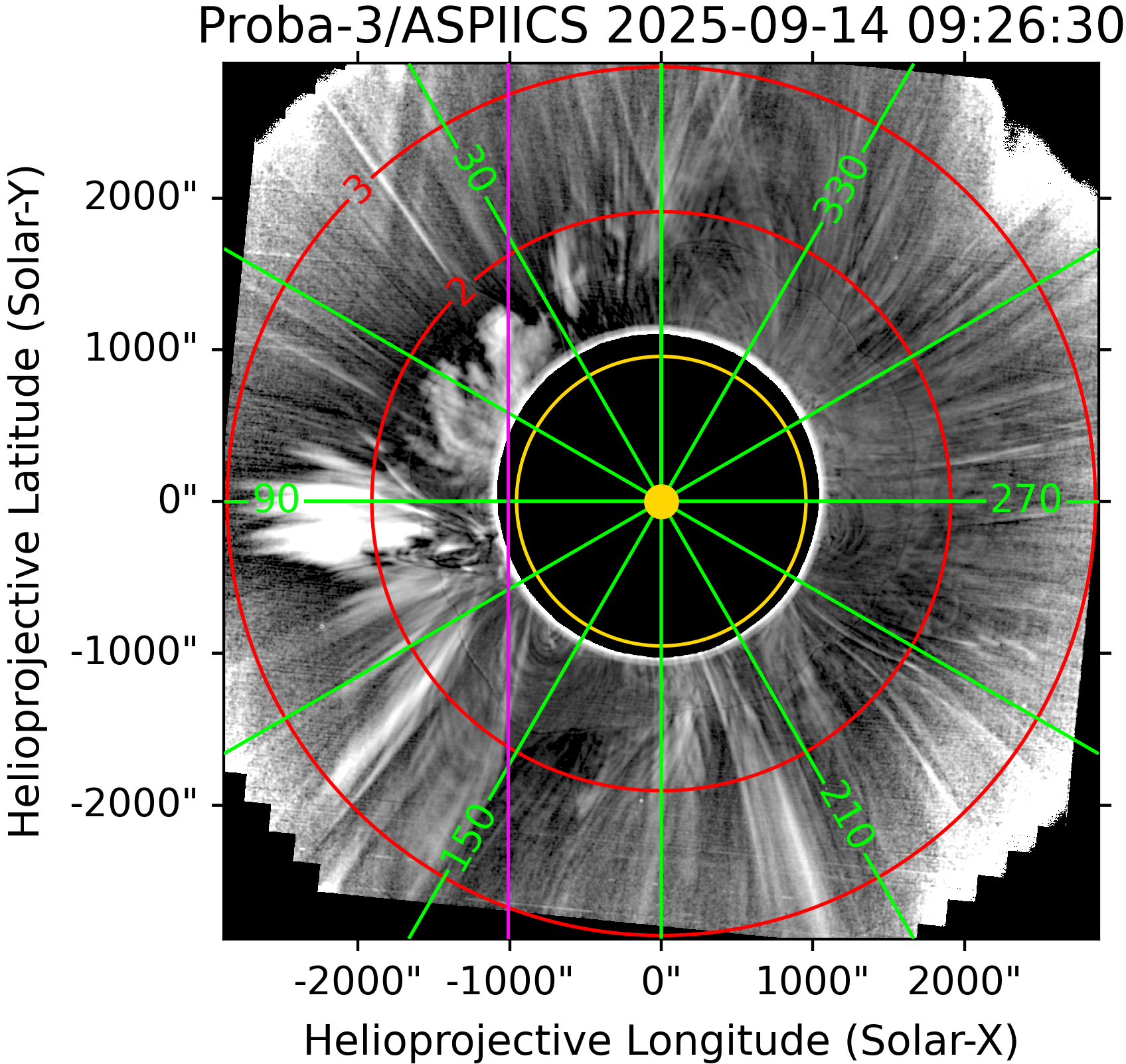}
\includegraphics[scale= 0.242,trim={0cm 0cm 0cm 0cm},clip]{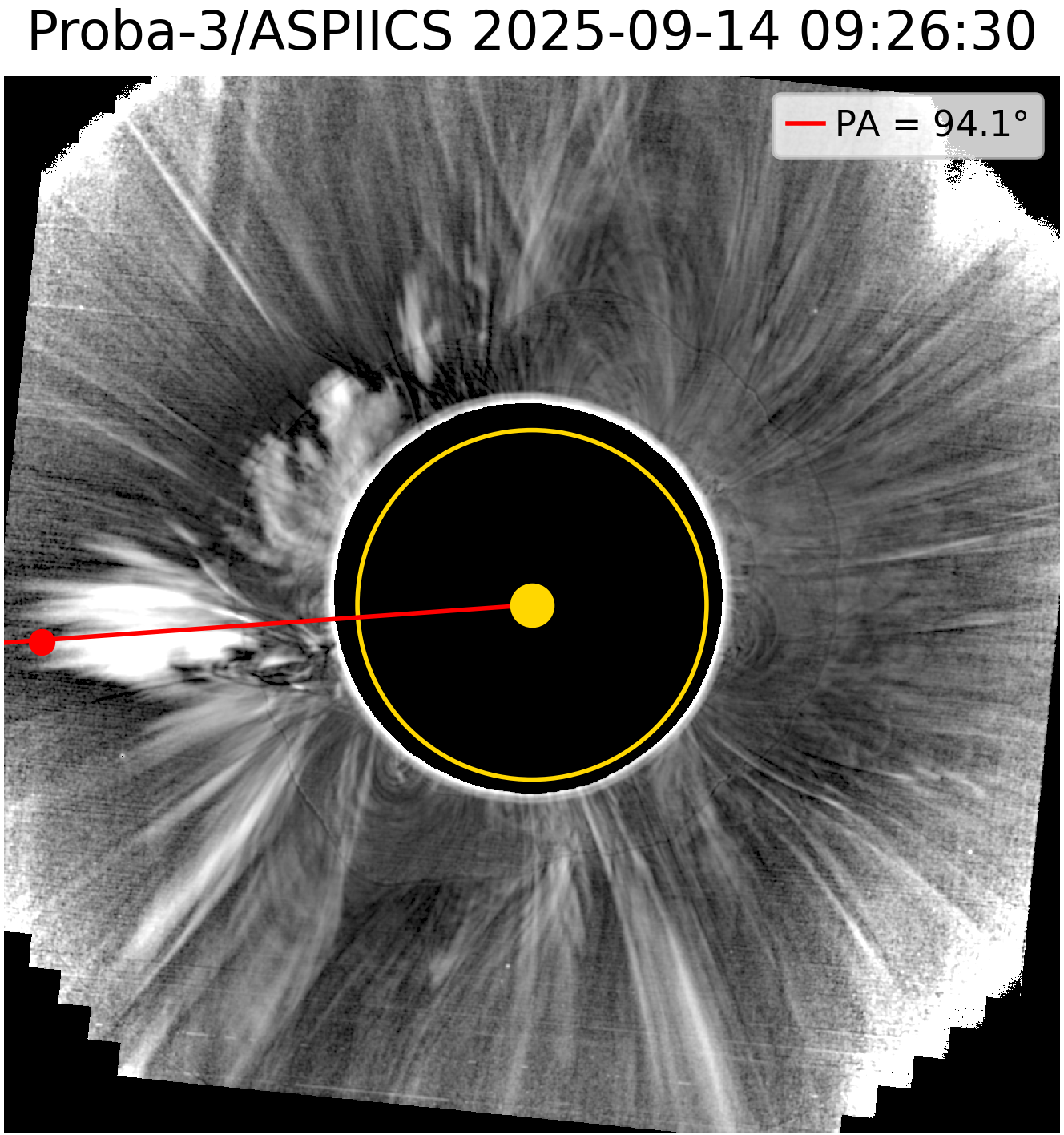}
\includegraphics[scale= 0.242,trim={0cm 0cm 0cm 0cm},clip]{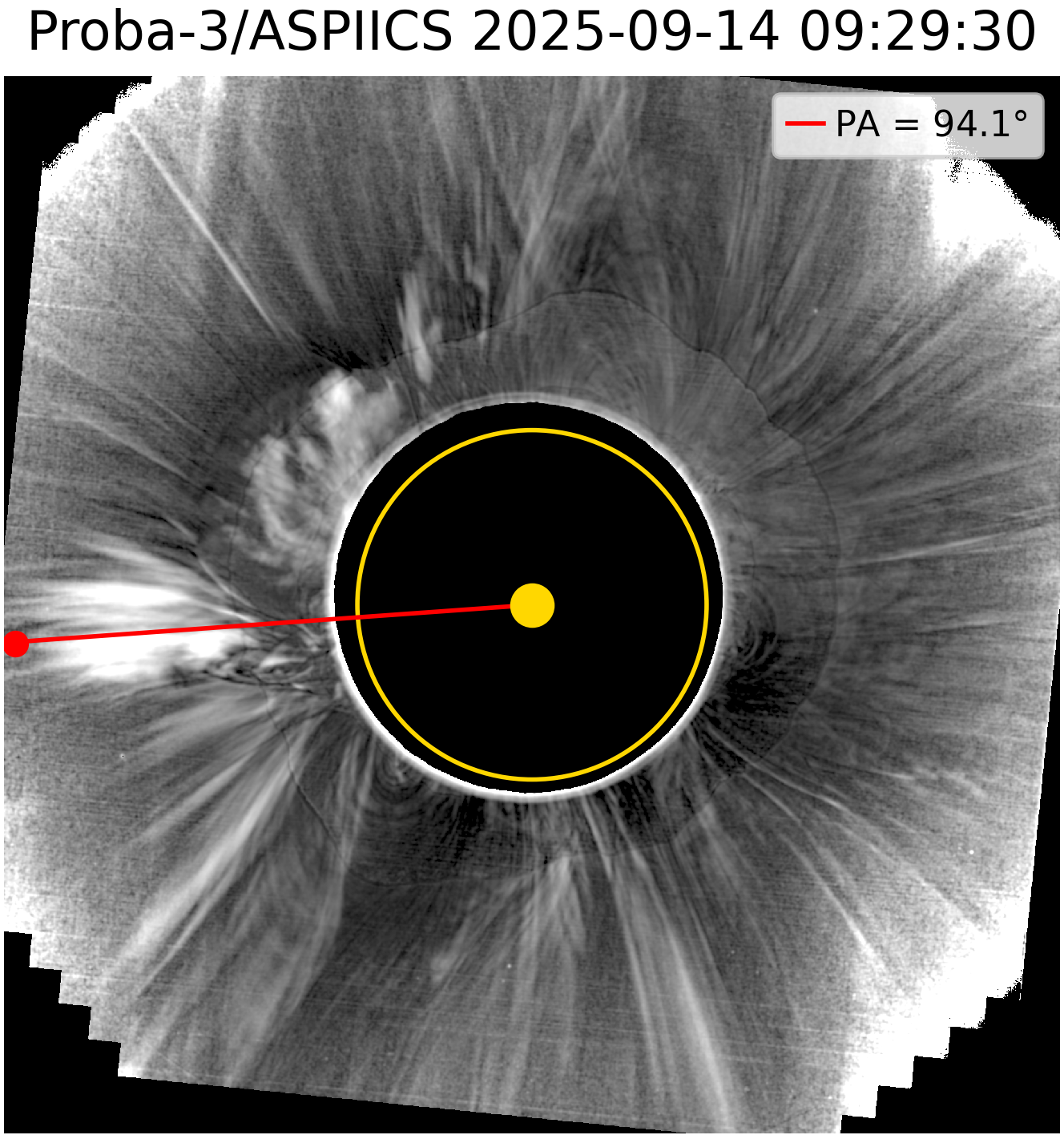}
\caption{Figure showing the ASPIICS wideband base-difference images of the 14 Sep CME. The left panel shows the position angles (green) along with the 2 and 3~$R_\odot$ contours (red). The solar limb is indicated by the yellow circle, and the solar center is marked by a yellow dot. The magenta line indicates the position of the VELC slit, whose center is located at a heliocentric distance of $\sim$1.06~$R_\odot$. The middle and right panels display the tracked CME heights at the initial and final observation times, respectively, along a position angle of 94.1$^\circ$.}
\label{fig:tracked_height_14_sep}
\end{figure*}

For the 14 Sep event, the left composite image shows slightly offset observation times between ASPIICS and SOHO/LASCO-C2, with the CME already partially present in the overlapping field of view. The CME structure below the occulting disk of LASCO-C2 in the low corona is clearly visible in the ASPIICS observations. The earliest available ASPIICS observation for this event is at approximately 09:22~UT, as no earlier wideband observations are available on that day. Moreover, a single Fe~XIV image is available at 09:23 UT; the lack of a pre-event image prevents the generation of base-difference images needed to isolate the CME signal. The middle and right panels show the CME progressively propagating beyond the ASPIICS FoV. Interestingly, the signature of this CME was also observed in an earlier phase of VELC observations at 08:56 UT, highlighting the importance of coordinated low-coronal observations.

For the 16 Sep event, the left composite image shows the CME within the LASCO-C2 occulter, as observed by ASPIICS at 04:10~UT. Unlike the previous event, this CME was captured from an earlier phase and was also detected in the Fe~XIV channel beginning at 03:54~UT, enabling direct comparison of CME kinematics derived from wideband and spectral-line observations in the low corona (Section~\ref{sec:cme_kinematics}). The middle and right panels illustrate the CME evolution within the overlapping FoV and its subsequent propagation beyond the ASPIICS FoV. Notably, the CME was detected even earlier in VELC observations at 03:56~UT, demonstrating synergy between ASPIICS and VELC for probing CME energetics and its evolution at low-coronal heights.

The source regions and onset times of both CMEs were identified using SDO/AIA 211~$\AA$ observations examined with JHelioviewer \citep{Mueller2010}. The 14 Sep CME originated near S15E90 at approximately 08:50~UT, while the 16 Sep CME originated near N05E55 at approximately 03:50~UT. However, the source locations are not clearly discernible in the EUV images, resulting in an uncertainty of about $\pm5^\circ$. The CME onset times are further supported by independent estimates from their kinematic evolution (Section~\ref{sec:cme_kinematics}) and by the initial flux enhancement observed in the VELC spectroscopic data (Section~\ref{sec:velc}). In the following section, we describe the methodology used to generate base-difference images from the ASPIICS wideband observations and to estimate the CME mass and number density for further estimation of CME energetics.

\begin{figure*}
\centering
\includegraphics[scale= 0.21,trim={0cm 0cm 0cm 0cm},clip]{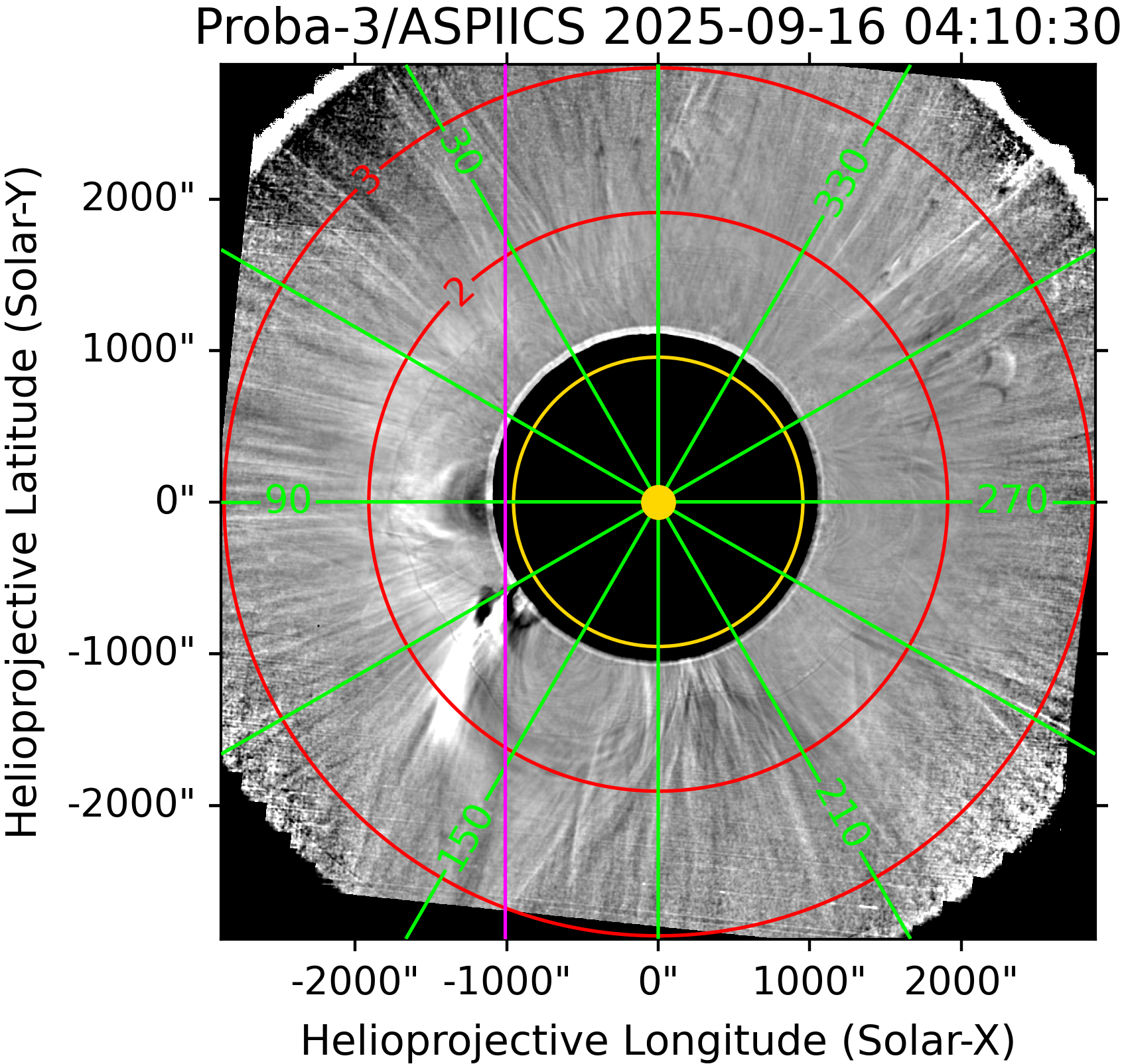}
\includegraphics[scale= 0.595,trim={0cm 0cm 0cm 0cm},clip]{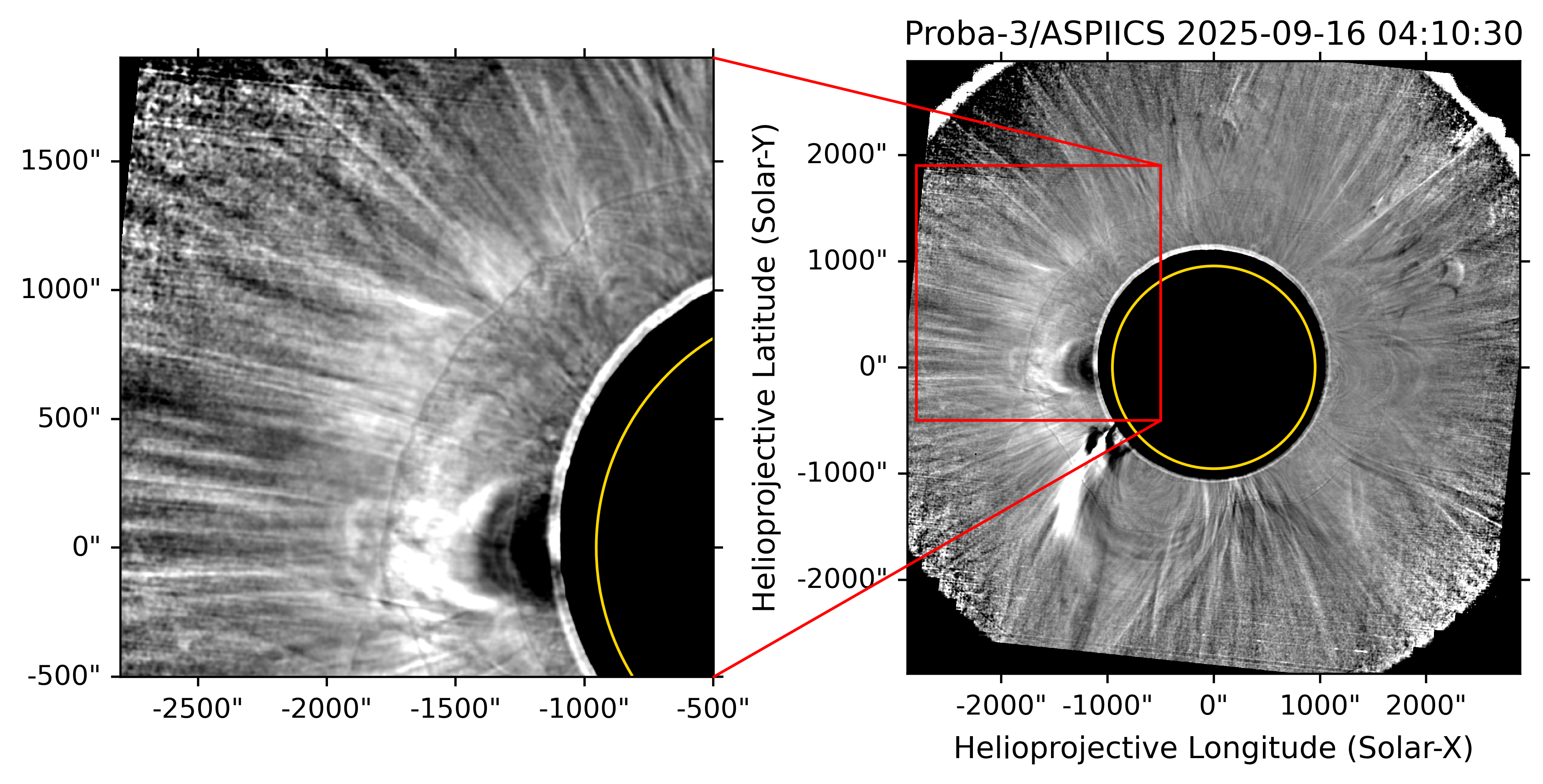}\\[0.3cm]
\includegraphics[scale= 0.19,trim={0cm 0cm 0cm 0cm},clip]{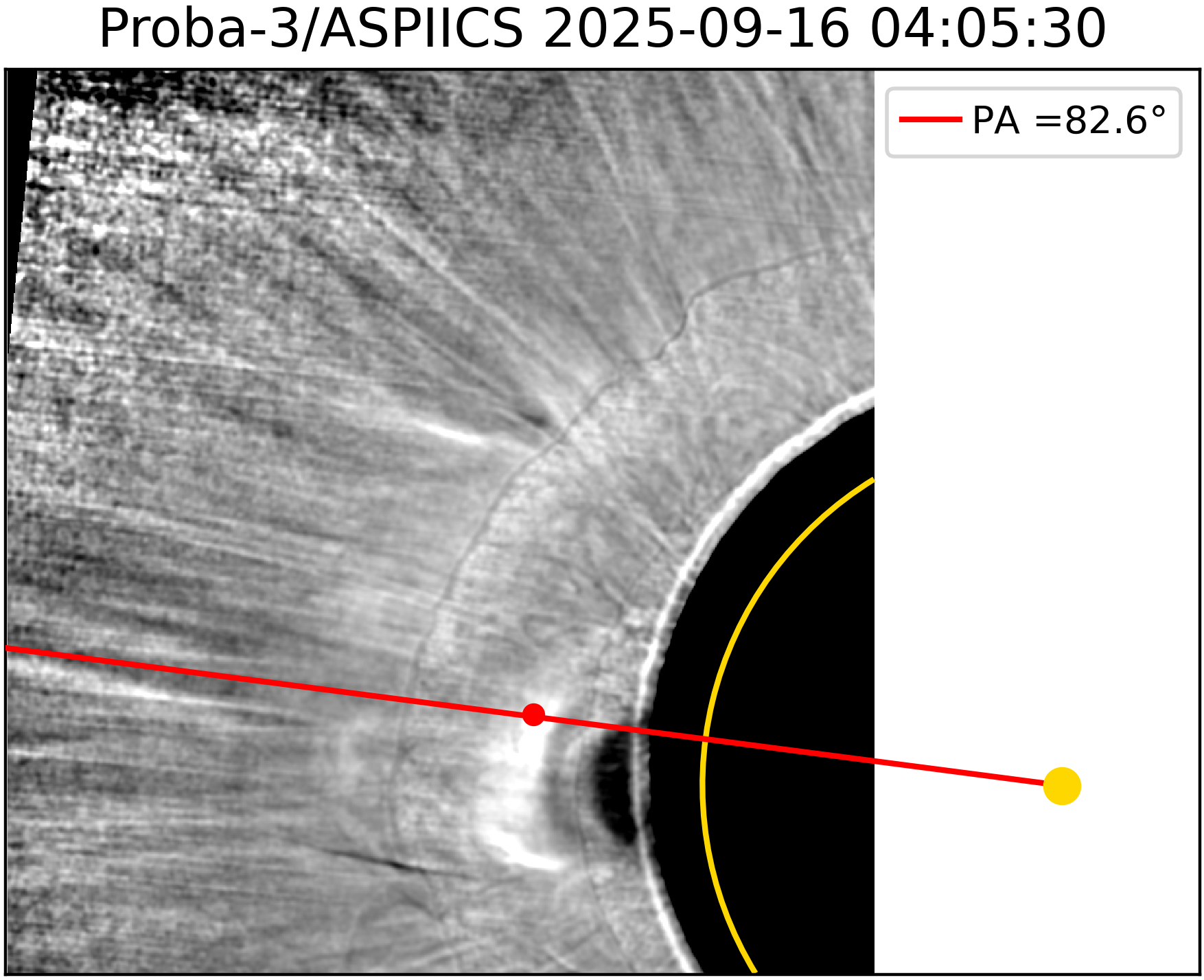}
\includegraphics[scale= 0.19,trim={0cm 0cm 0cm 0cm},clip]{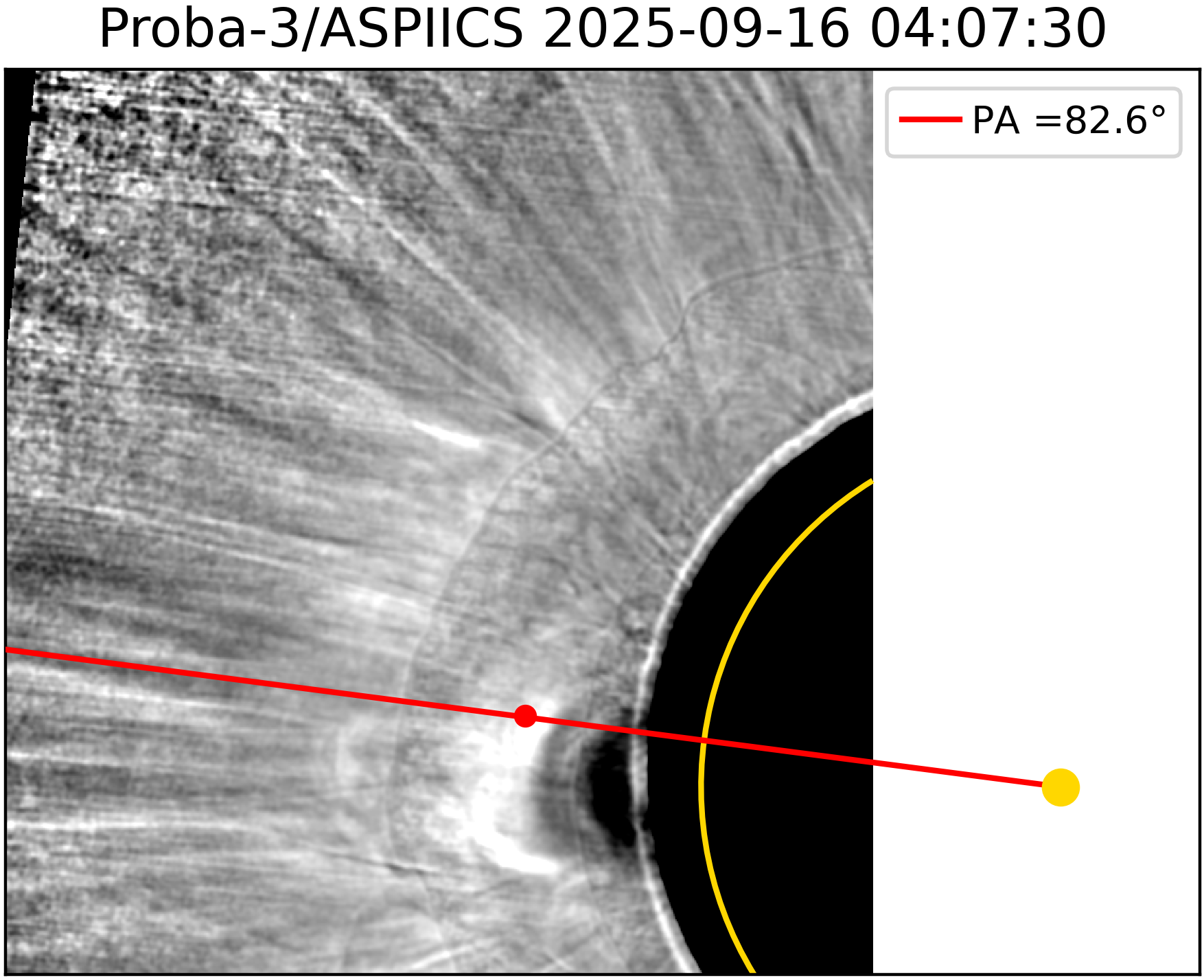}
\includegraphics[scale= 0.171,trim={0cm 0cm 0cm 0cm},clip]{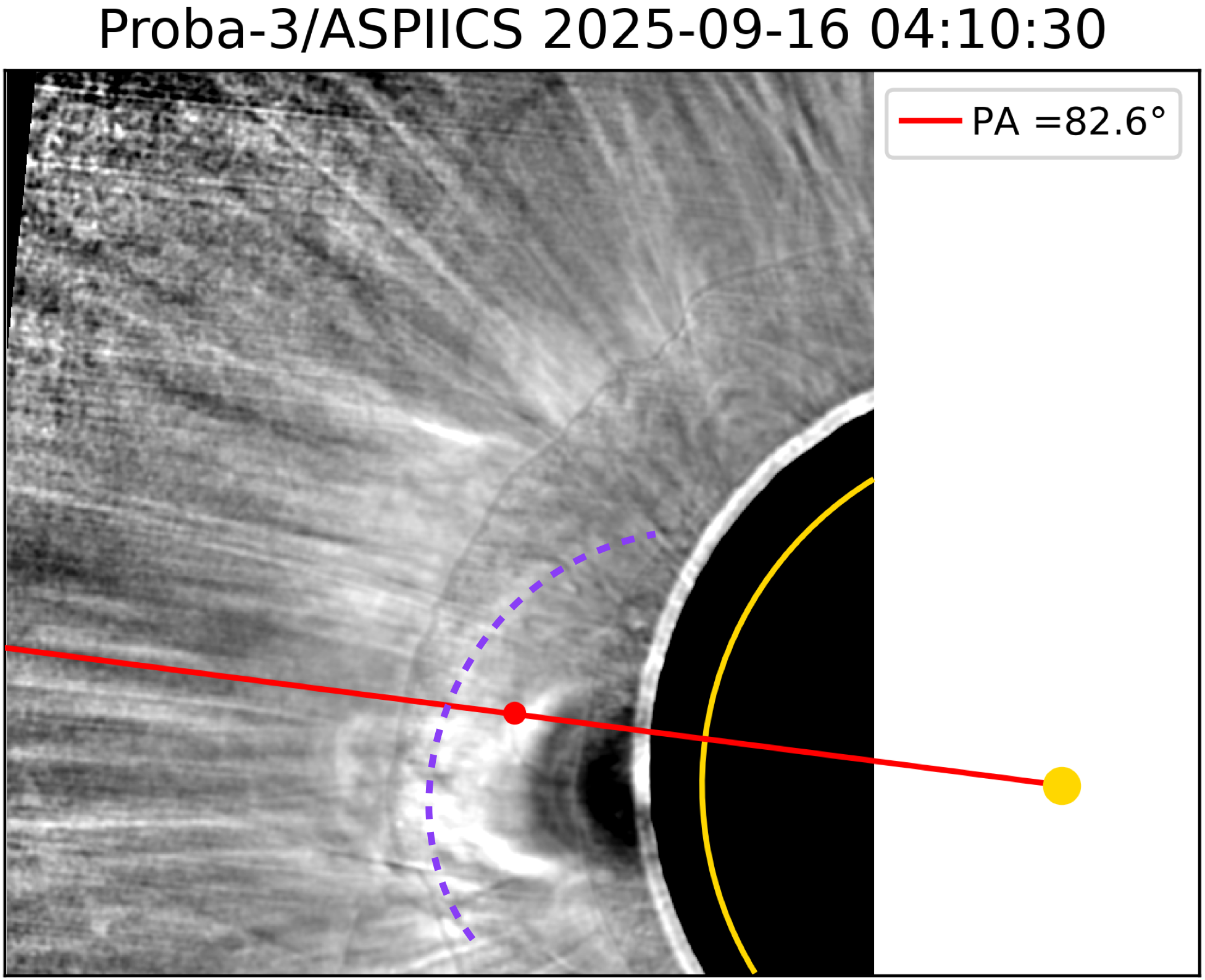}\\[0.3cm]
\includegraphics[scale= 0.19,trim={0cm 0cm 0cm 0cm},clip]{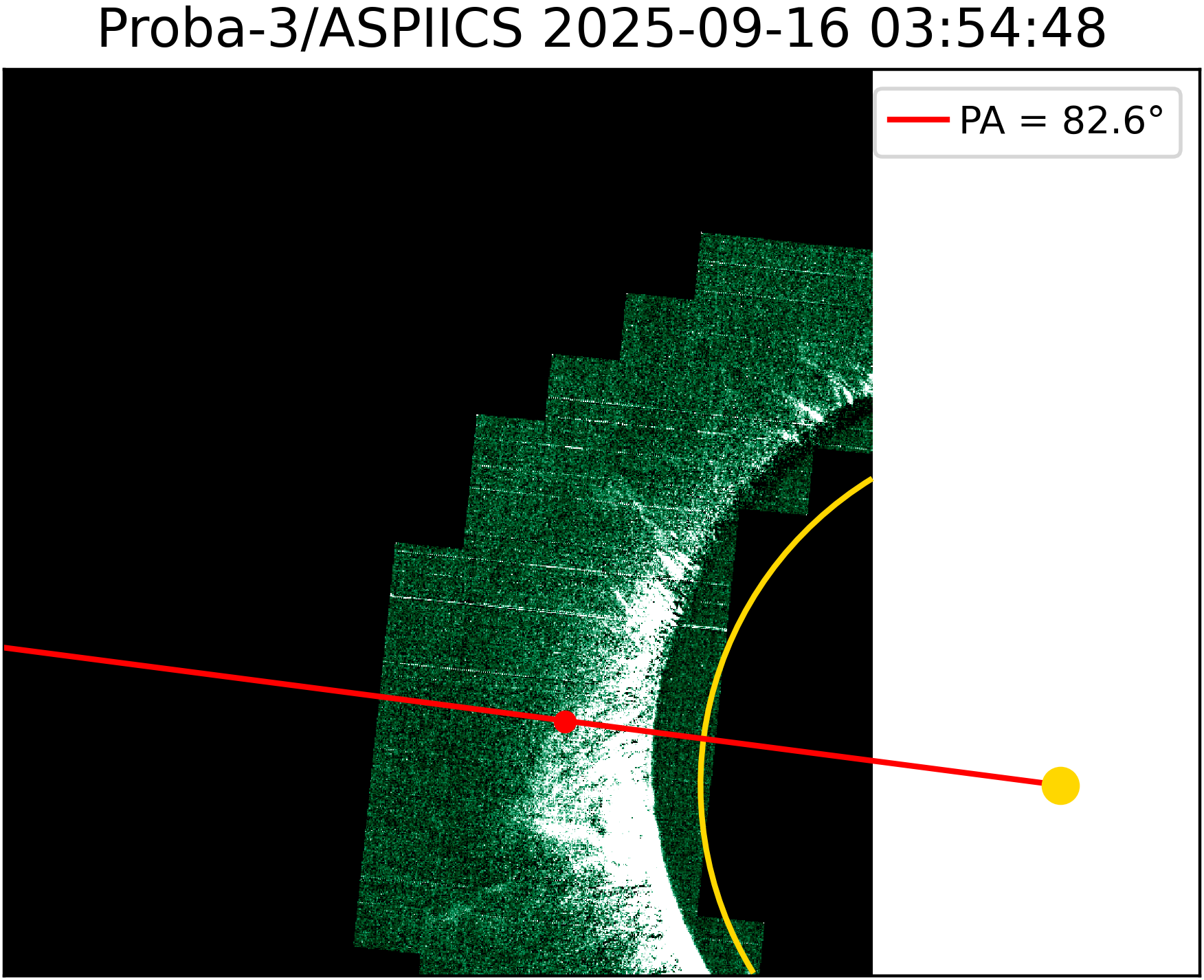}
\includegraphics[scale= 0.19,trim={0cm 0cm 0cm 0cm},clip]{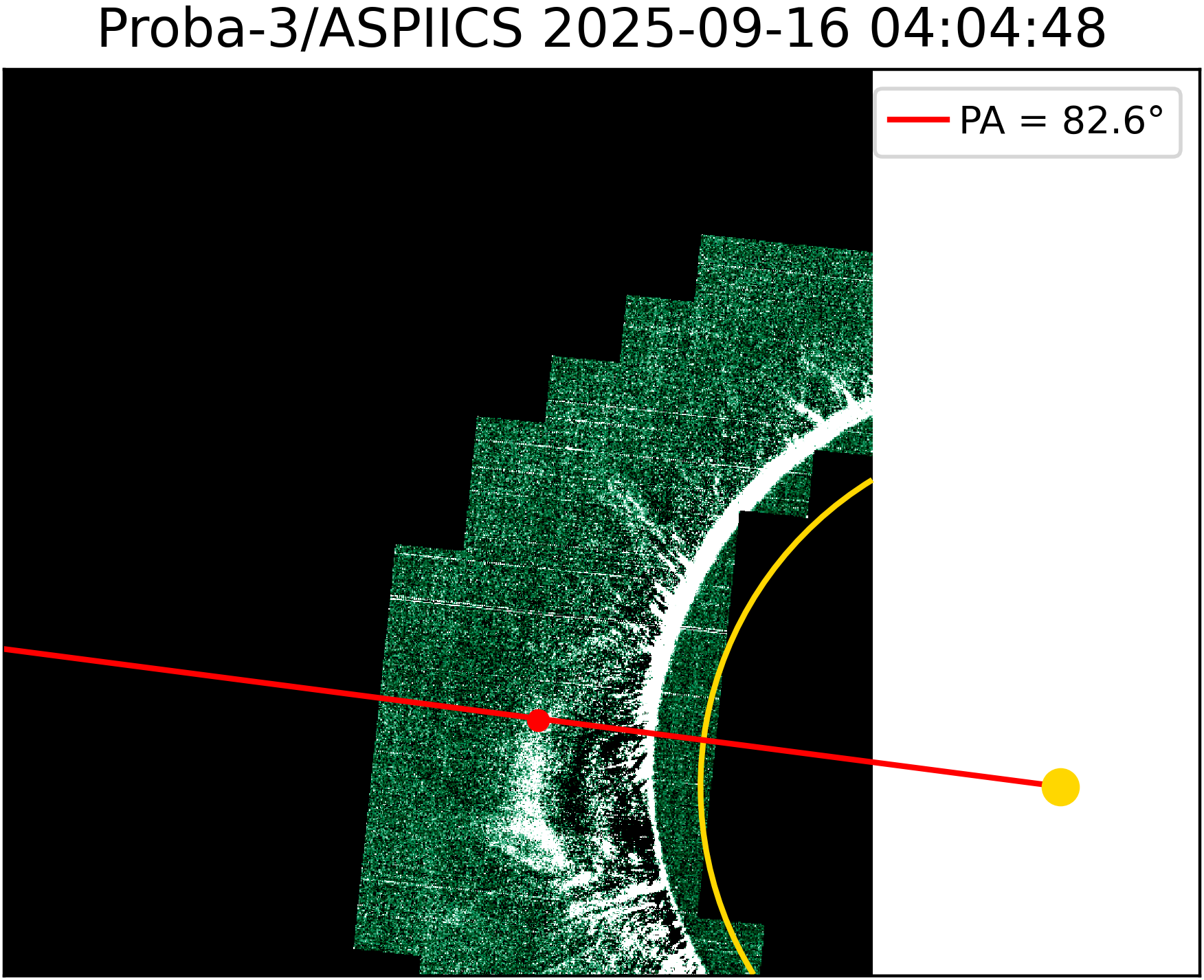}
\includegraphics[scale= 0.19,trim={0cm 0cm 0cm 0cm},clip]{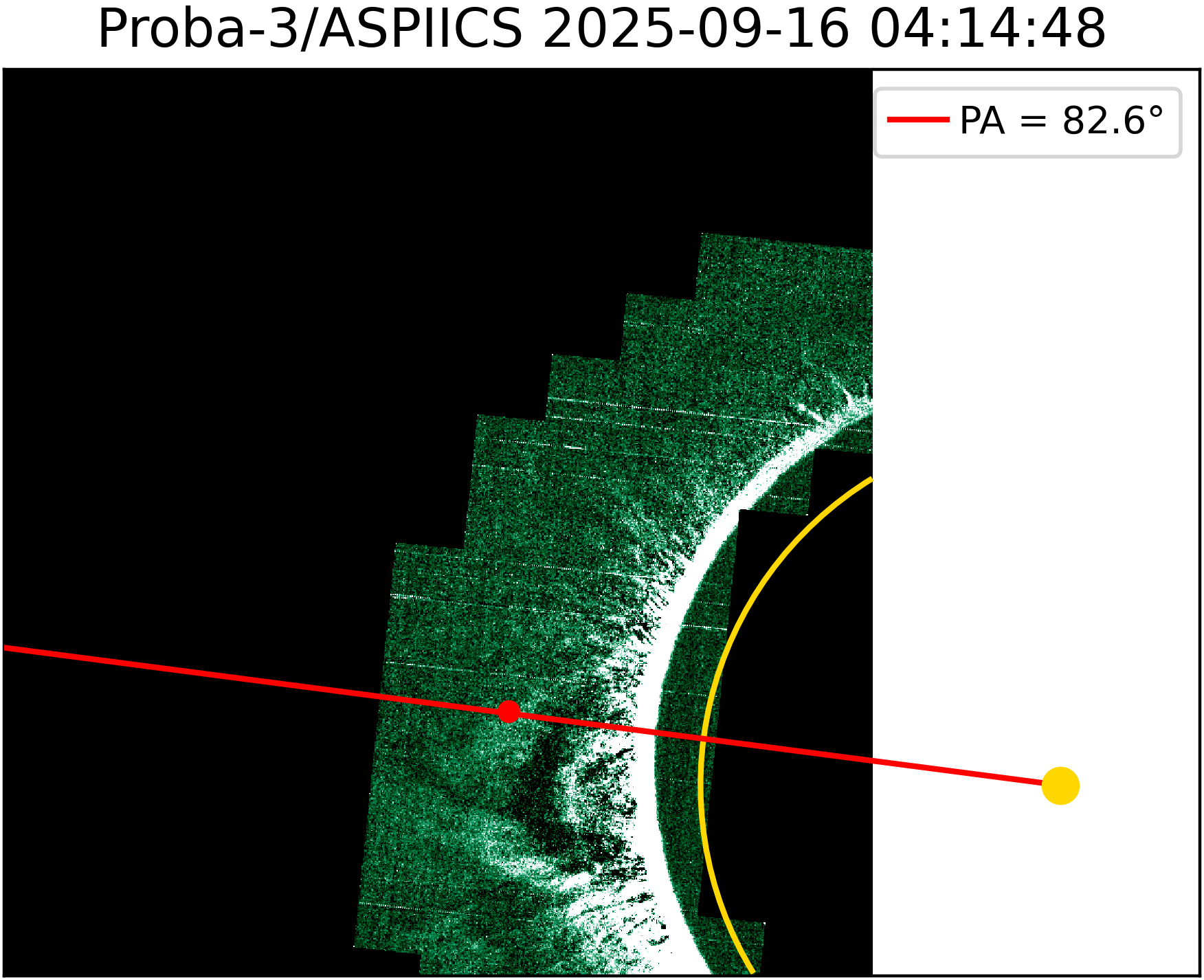}
\caption{Figure showing the ASPIICS wideband base-difference images of the 16 Sep CME. The left image in the top panel shows the position angles (green) and the 2 and 3~$R_\odot$ contours (red). The vertical magenta solid line indicates the position of the VELC slit, whose center is located at a heliocentric distance of $\sim$1.06~$R_\odot$. The solar limb and the Sun's center are marked by a yellow circle and a yellow dot, respectively. The right image shows a submap of the full field of view, highlighting the selected CME region. The magenta line indicates the position of the VELC slit, whose center is located at a heliocentric distance of $\sim$1.06~$R_\odot$. The middle panel presents the tracking of CME LE in red at multiple time steps. In the rightmost plot, the dashed purple curve indicates the presence of another wider feature, possibly associated with another CME (see Section~\ref{sec:limitstudy}). The bottom panel shows the tracking of the CME LE in the Fe~XIV channel at the initial, intermediate, and final observation times.}
\label{fig:tracked_height_16_sep}
\end{figure*}

\subsection{Kinematics, Mass, and Density Estimation of CMEs Using ASPIICS Observations} \label{sec:aspiics}

\subsubsection{Image Processing of ASPIICS} \label{sec:img_pro}

For the selected CMEs on 14 and 16 Sep, we use Level-3, version 02 data from the ASPIICS wideband channel, obtained from the P3SC ASPIICS data archive (\url{https://p3sc.oma.be/P3SC_archive/}). These Level-3, version 02 science-ready data are already corrected for instrumental effects, including image rotation to align Solar North upward and the removal of the background F-corona \citep{Zhukov2025}. To further isolate the CME emission and remove the background K-corona contribution, we constructed base-difference images. This involves generating a base image that represents the quiescent corona and subtracting it from the Level-3 coronagraphic images. The resulting base-difference images primarily represent the Thomson-scattered signal from the CME. The procedure used to construct the base-difference images is described below.

\begin{itemize}
\item The base (background) image is constructed by computing either the minimum or median intensity at each pixel from a set of images acquired during or prior to the CME. If the CME is already present in the earliest available ASPIICS observations, as is the case for the 14 Sep CME, the base image is generated by taking the minimum intensity over the full duration of the CME within the ASPIICS FoV. Otherwise, the base image is constructed by taking the median of images obtained approximately 10 minutes prior to the CME onset (pre-event), as was the case for the 16 Sep CME. Both approaches ensure that the resulting images primarily contain Thomson-scattered emission from the CME.

\item A standard isotropic Gaussian filter is applied to the base image to suppress high-frequency noise, using a normalized convolution approach. In this way, the Gaussian-weighted average is normalized by the contribution of valid neighboring pixels.

\item The processed base image is subtracted from the CME-containing images to remove the slowly varying background corona and isolate the CME emission. The resulting base-difference images are further smoothed using the same Gaussian filter to enhance the visibility of CME structures for reliable tracking. The Gaussian filtering introduces only minor changes in the total intensity ($<5\%$) of both the full image and the CME region, and negligible changes in the derived CME mass and number density ($<0.5\%$; Section~\ref{sec:cme_parameters}).
\end{itemize}

The left panels of Figures~\ref{fig:tracked_height_14_sep} and \ref{fig:tracked_height_16_sep} show the base-difference images of the 14 Sep and 16 Sep CMEs at 09:26:30 UT and 04:10:30 UT, respectively. In both images, the position angles (PAs) are marked in green, while red contours denote the 2 and 3~$R_\odot$ circles. To enhance CME visibility and facilitate identification of the leading edge (LE) and CME boundaries, each base-difference image is normalized by dividing the intensity of each pixel with the corresponding intensity in the base image. However, all CME physical parameters presented in Section~\ref{sec:cme_parameters} are derived from the unnormalized base-difference images. For the 16 Sep CME, we additionally analyze Fe~XIV Level-2 images with a 5-minute cadence. These images are rotated to align Solar North upward, and base-difference images are constructed using a pre-event base image.

\subsubsection{Kinematics of CMEs using ASPIICS} \label{sec:cme_kinematics}

\begin{table*}
\tiny
\renewcommand{\arraystretch}{1.9}
% \begin{tabular}{cccccccc}
\begin{tabular*}{\textwidth}{@{\extracolsep{\fill}}cccccccc}
\hline
& & & \multirow{2}{*}{Height} & \multirow{2}{*}{Speed} & \multirow{2}{*}{Mass} & \multirow{2}{*}{Volume} & {Number}\\

\multirow{3}{*}{Date} & \multirow{3}{*}{Instrument} & \multirow{3}{*}{Estimates} & \multirow{2}{*}{($R_\odot$)} & \multirow{2}{*}{($km~s^{-1}$)} &  \multirow{2}{*}{($10^{14}$ g)} & \multirow{2}{*}{($10^{30}~cm^{3}$)} & {density} \\

&  & &  &  &  &  & {($10^8~cm^{-3}$)}\\

& & & ($\sim$Error)  & ($\sim$Error) & ($\sim$Error) & ($\sim$Error) & {($\sim$Error)} \\
\hline

\multirow{3}{*}{14 Sep} & \multirow{2}{*}{ASPIICS} & Observed & {2.57 ($\pm0.026$)-2.97 ($\pm0.03$)} & {665 ($\pm10$)-620 ($\pm10$)} & 2.98 ($\pm$0.45) & {22 ($\pm$3.3)-55 ($\pm$8.3)} & {0.05 ($\pm0.011$)-0.03 ($\pm0.006$)} \\

& & Extrapolated & 1.09  & -- & 3 ($\pm$0.45) & 1.7 ($\pm$0.26) & {0.9 ($\pm$0.19)} \\

& VELC & Observed & 1.09 & -- & {3.6 ($\pm$0.65)} & {0.68 ($\pm$0.1)} & {2.7 ($\pm$0.24)}\\
\hline
\multirow{3}{*}{16 Sep} & \multirow{2}{*}{ASPIICS} & Observed & {1.43 ($\pm$0.014)-1.54 ($\pm$0.015)} & {200 ($\pm3$)-120 ($\pm1.8$)} & 1.03 ($\pm0.15$) & {1.9 ($\pm0.28$)-8.6 ($\pm1.3$)} & {0.14 ($\pm0.029$)-0.1 ($\pm0.021$)} \\

& & Extrapolated & 1.12 & -- & 1.08 ($\pm$0.16) & 0.91 ($\pm$0.14) & {0.61 ($\pm$0.13)} \\
& VELC & Observed & 1.12 &--& {1.6 ($\pm$0.29)} & {0.29 ($\pm$0.044)} & {2.8 ($\pm$0.28)}\\
\hline
\end{tabular*}
\caption{Physical parameters of the 14 Sep and 16 Sep CMEs derived from ASPIICS and VELC observations. For each CME, the first row summarizes the ASPIICS observations, including the observed height range, speed range (from the first to the last tracked height), average mass, and the corresponding ranges of volume and number density. The second row lists the mass, volume, and number density obtained by back-extrapolating the ASPIICS measurements to 1.09~$R_\odot$ for 14 Sep CME and 1.12~$R_\odot$ for 16 Sep CME. The third row presents the mass, volume, and number density derived from VELC observations at the corresponding heights. The associated uncertainties in all estimated parameters are given in round brackets.}
\label{tab:compar_aspiics_velc}
\end{table*}

To investigate the kinematics of the selected CMEs, we tracked CME features along specific PAs in successive images. The 14 Sep CME is a narrow event with an angular width of $\sim10^\circ$ and a lateral extent of $\sim$0.5 $R_\odot$, for which the leading feature was tracked along a PA of $\sim94.1^\circ$ during 09:22:34--09:29:30~UT. The corresponding height range is listed in the fourth column of Table~\ref{tab:compar_aspiics_velc}, together with an estimated uncertainty of $\sim$1\% in the height measurements, derived from repeated tracking. The high spatial resolution and high cadence of ASPIICS observations significantly reduce the uncertainty in height measurements compared to the $\sim$5--10\% uncertainties reported in earlier studies using SOHO/LASCO, STEREO/COR, and subjective tracking methods applied to simultaneous multi-coronagraph observations \citep{Thernisien2009,Agarwal2024,Khuntia2025}. The middle and right panels of Figure~\ref{fig:tracked_height_14_sep} show the CME LE at heights of 2.82~$R_\odot$ and 2.97~$R_\odot$, respectively.

\begin{figure*}
\centering
\includegraphics[scale= 0.655,trim={0cm 0cm 0cm 0cm},clip]{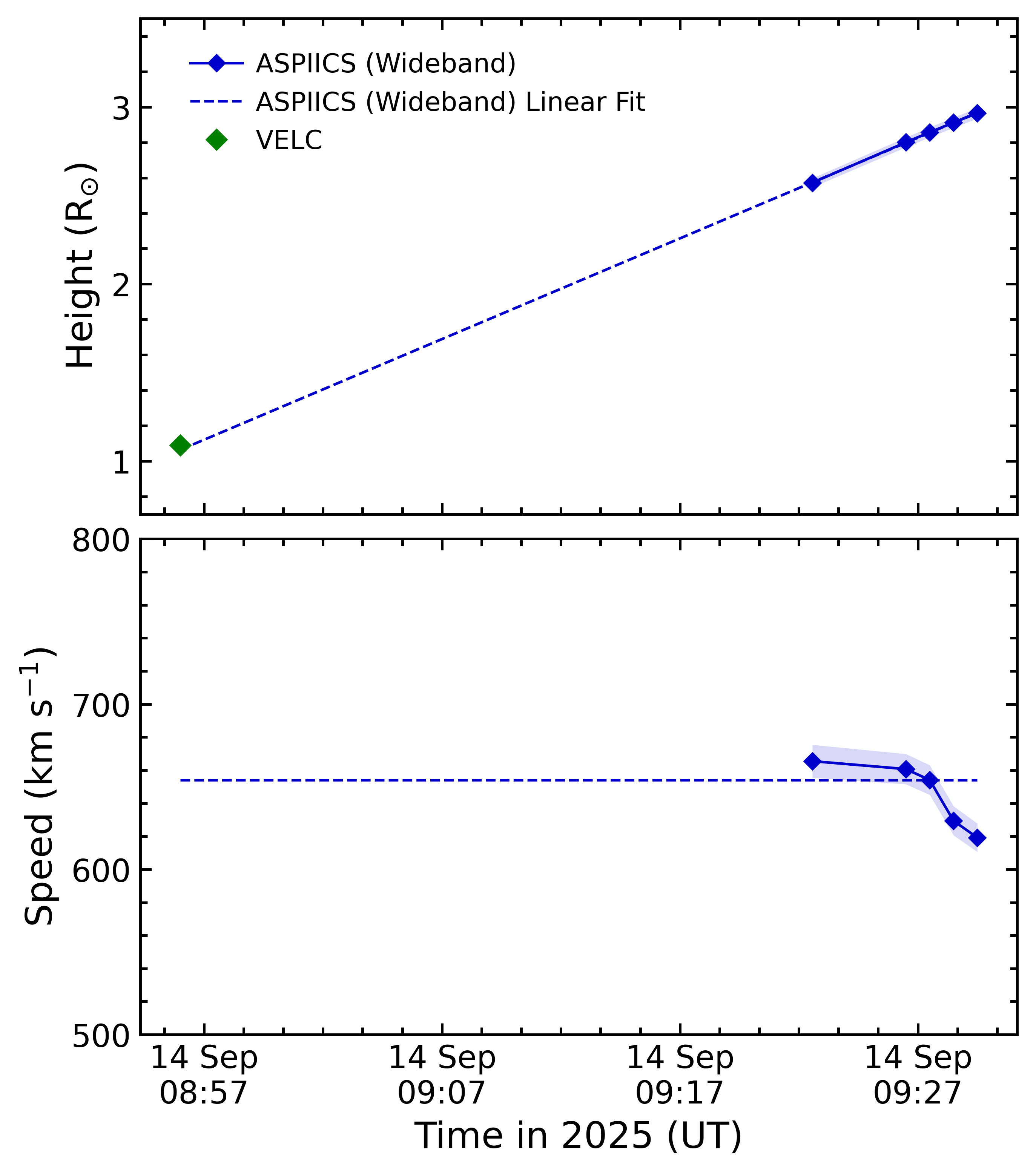}
\includegraphics[scale= 0.655,trim={0cm 0cm 0cm 0cm},clip]{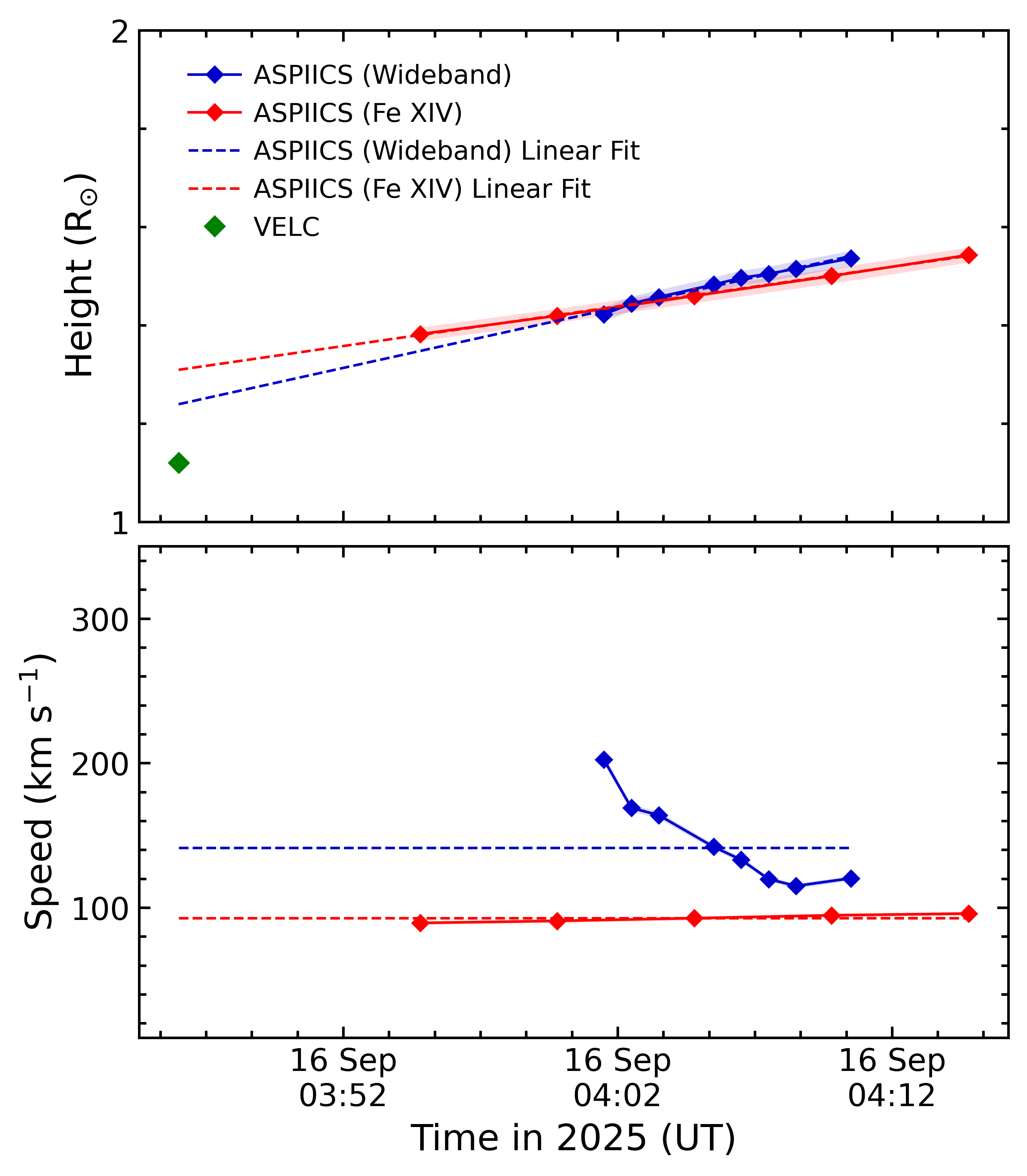}
\caption{The top panels show the height evolution of the CME LE, while the bottom panels show the corresponding LE speed for the 14 Sep CME in the left panel and the 16 Sep CME in the right panel. Solid lines represent the height measurements, while dashed lines show the corresponding linear fits. Blue and red curves correspond to the ASPIICS wideband and Fe~XIV observations, respectively. The green diamond marks the height of the enhanced-emission region at the onset time in the VELC observations. The corresponding CME speeds derived from the height--time measurements are shown in the bottom panels. The shaded regions represent the uncertainties in the height and speed estimates.}
\label{fig:kinematics}
\end{figure*}

The height--time measurements of 14 Sep CME derived from ASPIICS are shown by the blue solid line in the left panel of Figure~\ref{fig:kinematics}. The green diamond marks the height (1.09 R$_\odot$) of the enhanced emission region identified in the VELC observations at the onset time of the event (Section~\ref{sec:velc}). The blue dashed line represents a linear fit to the ASPIICS height--time measurements, back-extrapolated to the VELC onset time. The back-extrapolated CME height agrees well with the height of the enhanced-emission region observed by VELC, indicating consistency between the two independent observations. The corresponding LE speed was estimated using the moving-box linear-fit technique applied to the height--time measurements \citep{Agarwal2024} and is shown in the bottom panel of the figure. The associated uncertainties, estimated via error propagation, are shown as shaded regions. The ASPIICS measurements indicate a very slight decrease in CME speed over the observed height range, with the corresponding values listed in the fifth column of Table~\ref{tab:compar_aspiics_velc}.

The blue dashed line in the bottom panel represents the average CME speed derived from the linear fit to the ASPIICS height--time measurements, yielding a value of $\sim650$~km~s$^{-1}$. This is slightly higher than the linear-fit speed of 562~km~s$^{-1}$ reported in the CDAW catalog using LASCO-C2 and C3 observations. The difference is possible because the ASPIICS and LASCO measurements span different coronal height ranges, and the CME may have continued its gradual deceleration. Using the average speed from ASPIICS, the CME onset time is estimated to be $\sim$08{:}55~UT, in good agreement with the onset of enhanced flux observed in the VELC data (Section~\ref{sec:velc}).

On 16 Sep, two CMEs are observed nearly simultaneously at different PAs, as shown in the bottom panel of Figure~\ref{fig:composite_image}. The selected CME is relatively faint; therefore, a submap is used to enhance its visibility and better define its boundaries (right panel of Figure~\ref{fig:tracked_height_16_sep}). The CME first appears in ASPIICS observations at $\sim04{:}01{:}30$~UT, and its LE is tracked along a PA of $\sim82.6^\circ$, with tracked positions shown in red in the second panel of Figure~\ref{fig:tracked_height_16_sep}. The corresponding height range is listed in the fourth column of Table~\ref{tab:compar_aspiics_velc}, with the last reliable LE detection at 1.54~$R_\odot$ on 04:10:30~UT. The CME has an angular width of $\sim20^\circ$ and a lateral extent of $\sim0.5~R_\odot$ at its first tracked height, comparable to that of the 14 Sep CME despite being observed at a lower heliocentric distance.

Beyond this stage, the LE becomes diffuse and difficult to identify, due to the presence of a fast and wide propagating structure ahead of the CME; this structure is discussed further in detail in Section~\ref{sec:discussion}. The third panel of Figure~\ref{fig:tracked_height_16_sep} shows the CME LE identified in the Fe~XIV channel along the PA of $\sim82.6^\circ$, same as in the broad white-light band. In the Fe~XIV observations, the CME has a lateral extent of $\sim0.32~R_\odot$ at its first tracked height of $\sim1.38~R_\odot$. This lateral extent can be directly compared with the range of enhanced emission observed over the PAs extent in the VELC observations, as discussed in Section~\ref{sec:velc}.

The top panel of the right panel of Figure~\ref{fig:kinematics} shows the height--time measurements derived from the ASPIICS wideband in a blue solid line and Fe~XIV observations in a red solid line. The green diamond marks the height (1.12 R$_\odot$) of the enhanced emission region identified in the VELC observations at the onset time of the event (Section~\ref{sec:velc}). The blue and red dashed line represents a linear fit to the ASPIICS wideband and Fe~XIV height--time measurements, back-extrapolated to the VELC onset time. The back-extrapolated CME heights from wideband and Fe~XIV are slightly higher than the height of the enhanced-emission region observed by VELC. The CME LE speeds derived from the ASPIICS wideband and Fe~XIV observations are shown in the bottom panel of the figure as blue and red solid lines, respectively. In the wideband observations, the CME speed decreases at lower heights before approaching a nearly constant value, with the corresponding range listed in the fifth column of Table~\ref{tab:compar_aspiics_velc}.

In the bottom panel of the figure, the blue and red dashed lines represent the average CME LE speed derived from the linear fit to the ASPIICS wideband and Fe~XIV observations, respectively. The average LE speed from ASPIICS wideband is $\sim$145~km~s$^{-1}$, comparable to the value of 166~km~s$^{-1}$ reported in the CDAW catalog. Despite indications of deceleration in ASPIICS, possible differences in the CME features tracked, and the coronal height ranges sampled by ASPIICS and LASCO complicate a direct comparison of the measured speeds. In contrast, the Fe~XIV LE speed remains nearly constant at $\sim$90~km~s$^{-1}$. The difference likely arises because white-light observations trace the bulk electron density distribution of the CME, whereas Fe~XIV emission selectively samples plasma within a restricted range of temperature and ionization conditions. Consequently, the features identified in the two channels may trace different plasma components and need not exhibit identical speeds. Using the average wideband speed, the CME onset time is estimated to be $\sim03{:}30$~UT, earlier than the onset of enhanced flux observed in the VELC data (Section~\ref{sec:velc}), likely due to the assumption of constant propagation speed despite the observed deceleration at lower coronal heights. In the next section, we describe the methodology for estimating the CME number density and mass using Thomson scattering theory \citep{Billings1966,Howard2009}.

\subsubsection{Estimation of CME Mass, Volume, and Number Density using ASPIICS} \label{sec:cme_parameters}

Using the white-light coronagraphic observations, the total number of electrons contributing to the observed scattered intensity is estimated by dividing the excess observed brightness associated with the CME ($B_{\rm obs}$) by the brightness of a single electron ($B_e(\theta)$), computed from the Thomson scattering function \citep{Billings1966,Vourlidas2000,Howard2009}. Here, $B_e(\theta)$ represents the brightness produced by a single electron located at an angle $\theta$ from the plane of the sky (POS). Since the observed white-light brightness is the sum of the Thomson-scattered emission from all electrons along the line of sight, the ratio $B_{\rm obs}/B_e(\theta)$ provides an estimate of the total number of electrons in the CME. The estimated number of electrons is then converted into CME mass by assuming a plasma composition of 90\% hydrogen and 10\% helium, corresponding to a mass of $1.97\times10^{-24}$~g per electron, yielding the total CME mass as $M = N_e \times 1.97\times10^{-24}$~g \citep{Billings1966,Vourlidas2000,Howard2009,Mishra2014}. The mass estimates are corrected for the CME propagation angle relative to the POS, adopting POS angles of 0$^\circ$ and 35$^\circ$ for the 14 and 16 Sep CMEs, respectively. However, the estimated mass remains subject to uncertainties arising from the unknown CME depth and density distribution along the line of sight, which can lead to an underestimation of the CME mass even after the propagation-angle correction \citep{Colaninno2009}. Other sources of uncertainty are discussed in Section~\ref{sec:cmkimade}.

\begin{figure*}
\centering
\includegraphics[scale= 0.24,trim={0cm 0cm 0cm 0cm},clip]{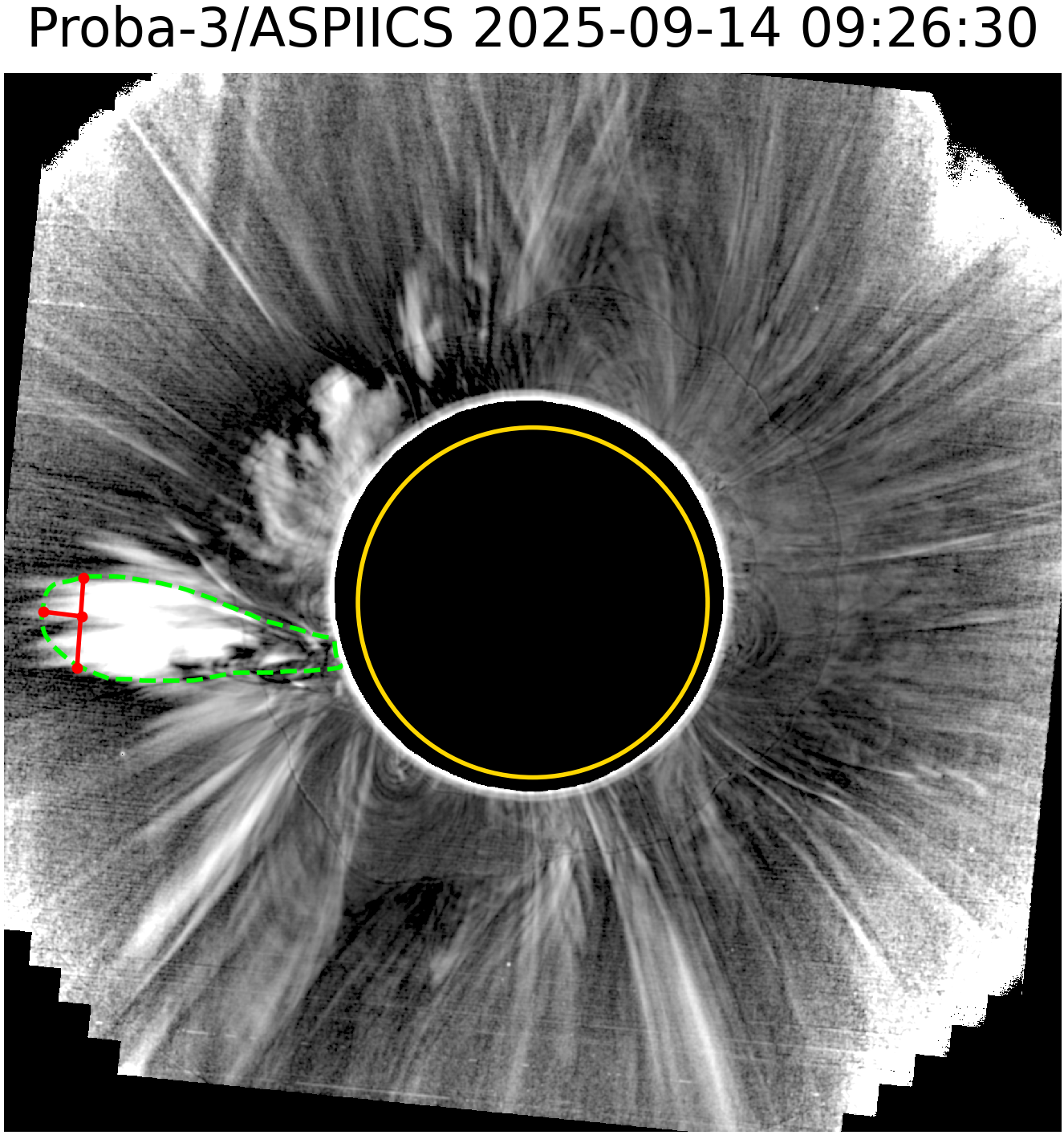}
\includegraphics[scale= 0.24,trim={0cm 0cm 0cm 0cm},clip]{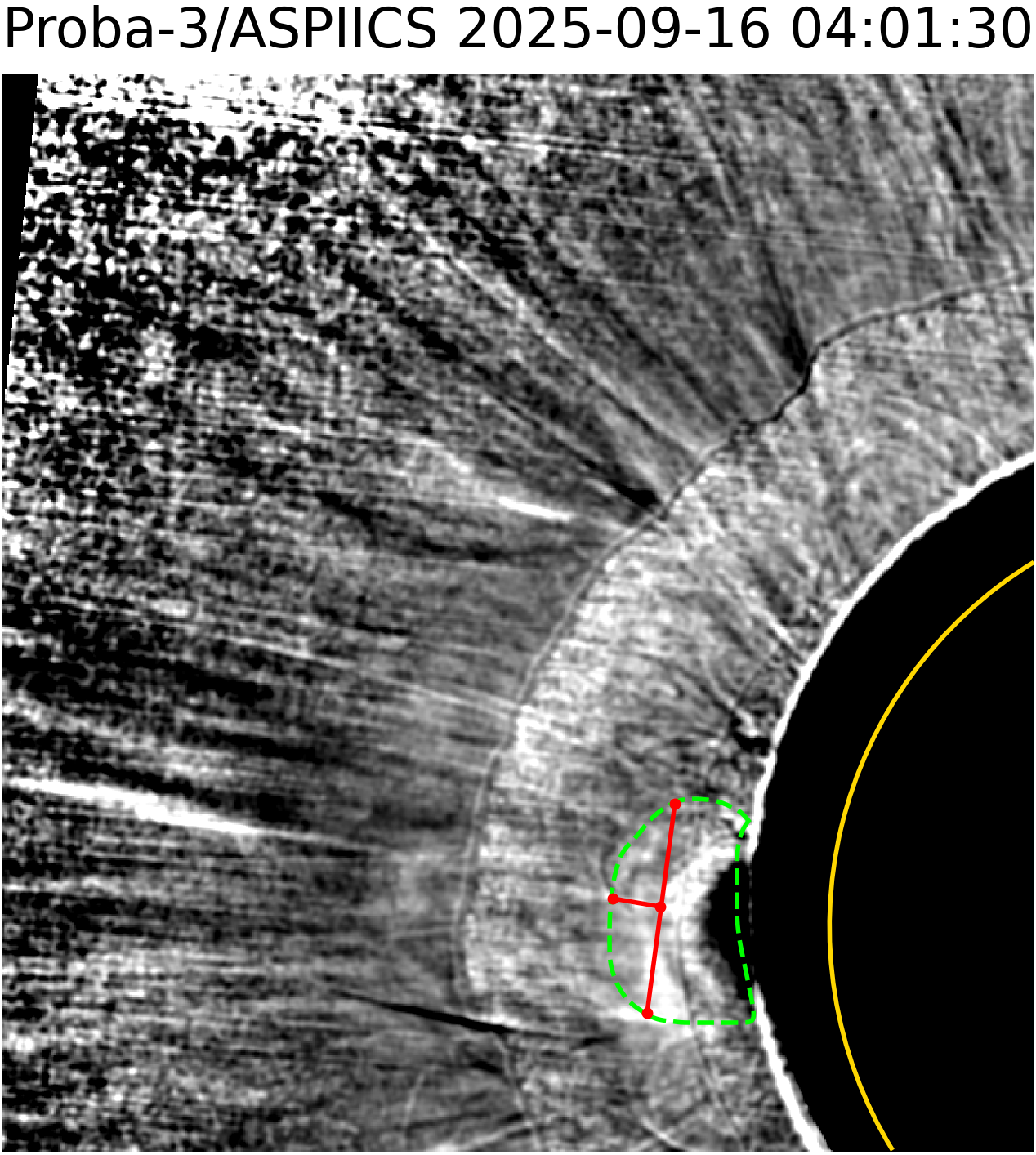}
\includegraphics[scale= 0.21,trim={0cm 0cm 0cm 0cm},clip]{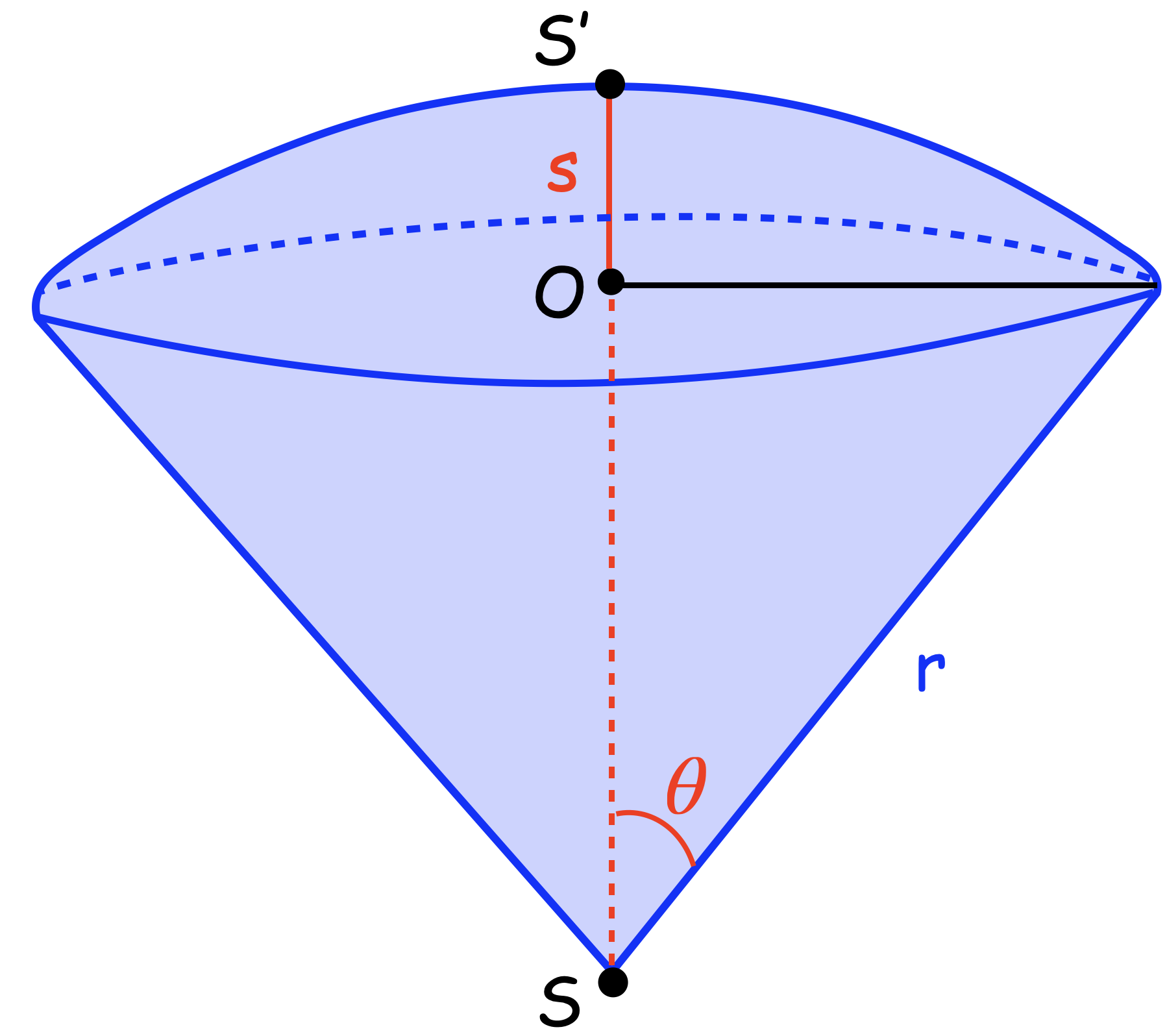}
\caption{The left panel shows the selected CME boundary (green contour) used to estimate the mass of the 14 Sep CME, while the middle panel shows the corresponding boundary for the 16 Sep CME. The right panel illustrates the spherical-sector geometry adopted for estimating the CME volume. In the left and middle panels, the vertical red lines indicate the lateral extent of the CME, whereas the horizontal red lines represent the spherical-sector height.}
\label{fig:mass_image}
\end{figure*}

The left and middle panels of Figure~\ref{fig:mass_image} show representative green contours used to define the CME boundaries for the 14 and 16 Sep CMEs, respectively. Similar contours were selected at all tracked heights to estimate the CME mass from the enclosed intensity. To quantify the uncertainty associated with the contour selection, the CME boundaries were repeatedly traced using slightly different contours. The resulting spread in the derived masses was typically $\sim$10--20\% at a given height for both events. Accordingly, we adopt a representative uncertainty of 15\% associated with the contour selection for the reported CME mass and propagate this uncertainty to the density estimates. 

Using this approach, the estimated number of electrons for the 14 Sep CME, ranges from $(1.14\pm0.17)\times10^{38}$ to $(1.79\pm0.27)\times10^{38}$, corresponding to a mass range of $(2.2\pm0.3)\times10^{14}$ to $(3.5\pm0.5)\times10^{14}$~g over a height range of $\sim$2.57--2.97~$R_\odot$. For the 16 Sep CME, the number of electrons ranges from $(2.67\pm0.4)\times10^{37}$ to $(7.9\pm1.2)\times10^{37}$, corresponding to a mass range of $(5.3\pm0.8)\times10^{13}$ to $(1.6\pm0.2)\times10^{14}$~g over a height range of $\sim$1.43--1.49~$R_\odot$. The corresponding average values of mass are listed in the sixth column of Table~\ref{tab:compar_aspiics_velc}. The estimated number of electrons and CME masses are broadly consistent with the values reported in earlier studies at comparable coronal heights \citep{Ramesh2000}. For both events, the CME mass increases with leading-edge height, consistent with previous studies \citep{Bein2013}, with a more pronounced increase observed for the 16 Sep CME.

For the 16 Sep CME, beyond heights of $\sim$1.49~$R_\odot$, the CME becomes diffuse, and its boundaries can no longer be reliably identified, although the LE feature remains trackable along the PA of $82.6^\circ$ up to $\sim$1.54~$R_\odot$ (Section~\ref{sec:cme_kinematics}). Therefore, the CME mass is assumed to remain constant beyond the last height ($\sim$1.49~$R_\odot$) at which the boundary is clearly identifiable to estimate the kinetic energy evolution at these larger heights (Section~\ref{sec:comparing_energy}).

To estimate the CME volume, we adopt a spherical-sector geometry, illustrated in the right panel of Figure~\ref{fig:mass_image}, which provides a more realistic representation of CME structure than a full-cone approximation \citep{Bein2013}. The volume of a spherical sector is given by $V=\frac{2}{3}\pi r^3(1-\cos\theta) \Rightarrow V \propto r^3$, where $r$ is the heliocentric distance and $\theta$ is the half angular width of the CME. Using the CME projected area ($A = 2\pi rs$) and sector height ($s=r(1-cos\theta)$), marked as $\mathrm{OS'}$ in the figure. Substituting these expressions for A and s into the spherical-sector volume formula yields $V=A^2/(6\pi s)$. This method is applied over the height range where the CME boundary is clearly identifiable, i.e., throughout the tracked height range for the 14 Sep CME and over $\sim$1.43--1.49~$R_\odot$ for the 16 Sep CME.

Beyond this range, the 16 Sep CME becomes diffuse, and its volume is extrapolated from the last reliably estimated value using the relation $V \propto r^3$. The same relation is also used to back-extrapolate the volumes of the 14 and 16 Sep CMEs from their first tracked heights to 1.09~$R_\odot$ and 1.12~$R_\odot$, respectively, for comparison with the VELC observations (Section~\ref{sec:velc}).

At the first tracked height, the values of sector height ($s$) measured from the images are 0.23 $R_\odot$ and 0.14~$R_\odot$ for the 14 and 16 Sep CMEs, respectively. The CME projected area at each tracked height is estimated by identifying the CME boundary. A representative boundary, outlined by the green contour in the left and middle panels of Figure~\ref{fig:mass_image}, is shown for the 14 and 16 Sep CMEs, respectively. The projected area is then calculated by counting the pixels enclosed within the contour and converting the corresponding pixel area into physical units \citep{Mishra2014}. The estimated CME areas at the first tracked height are $\sim2.6\times10^{21}$~cm$^2$ and $\sim5.9\times10^{20}$~cm$^2$ for the 14 and 16 Sep CMEs, respectively. Using these area estimates together with the corresponding number of electrons, we estimate column electron densities of $\sim4.2\times10^{16}$~cm$^{-2}$ for 14 Sep CME and $\sim4.5\times10^{16}$~cm$^{-2}$ for 16 Sep CME.

The estimated areas correspond to volumes of $\sim2.2\times10^{31}$~cm$^3$ and $\sim1.9\times10^{30}$~cm$^3$ for the 14 and 16 Sep CME, respectively. The range of estimated volumes for both events at their tracked height range is listed in the seventh column of the Table~\ref{tab:compar_aspiics_velc}. The uncertainty in the CME volume was estimated by repeatedly measuring the projected area and sector height, and using the resulting variation in the calculated CME volume, yielding an uncertainty of $\sim15\%$. The estimated volume of the 14 Sep CME exceeds that of the 16 Sep CME because the former is observed at higher heights, where its projected area is larger. Using the estimated volume and total number of electrons, the electron number density is estimated at each tracked height. The resulting ranges of volume and number density, together with their uncertainties from error propagation, are listed in the seventh and eighth columns of Table~\ref{tab:compar_aspiics_velc}.

\begin{figure*}
\centering
\includegraphics[scale= 0.73,trim={0cm 0cm 0cm 0cm},clip]{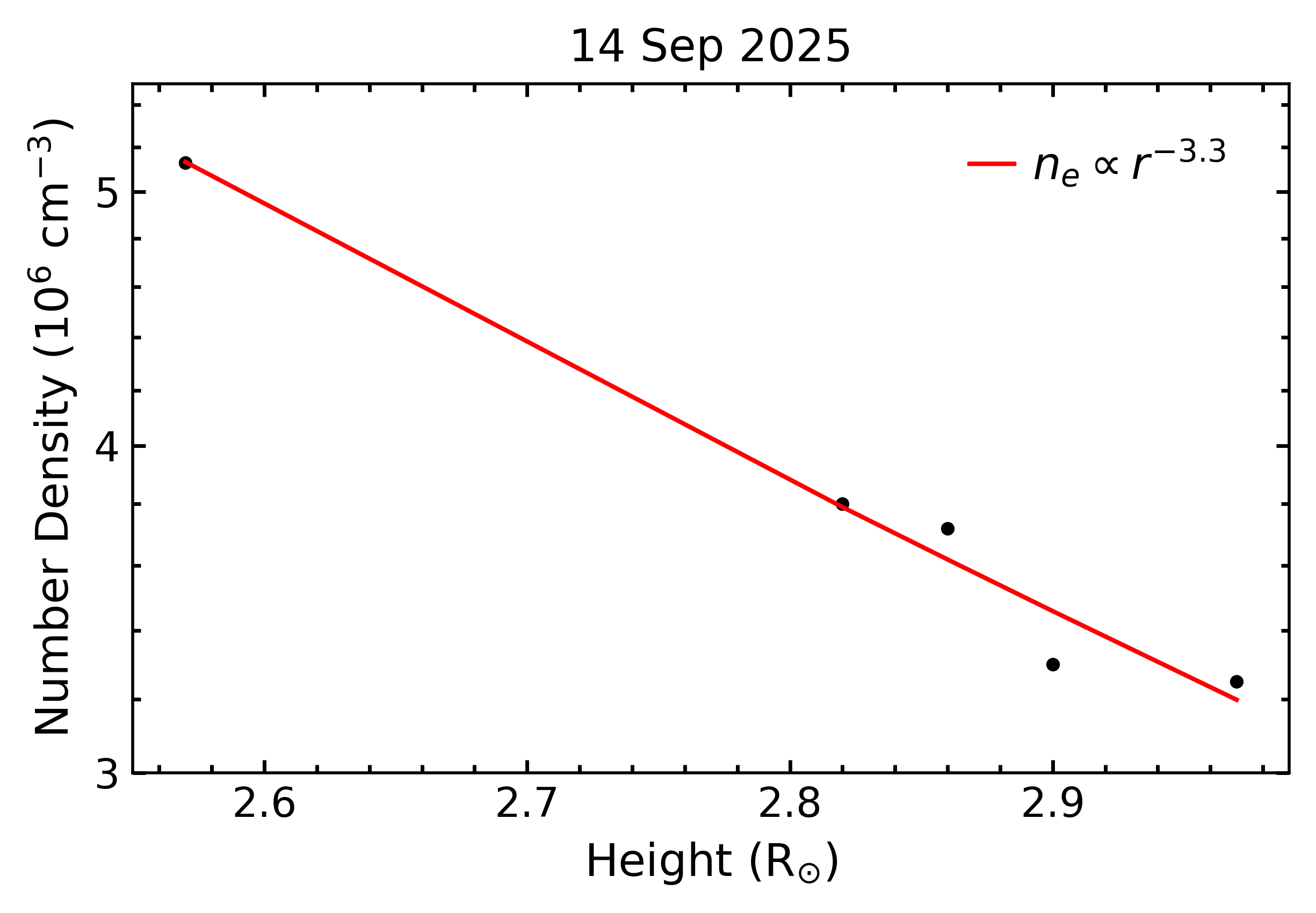}
\includegraphics[scale= 0.73,trim={0cm 0cm 0cm 0cm},clip]{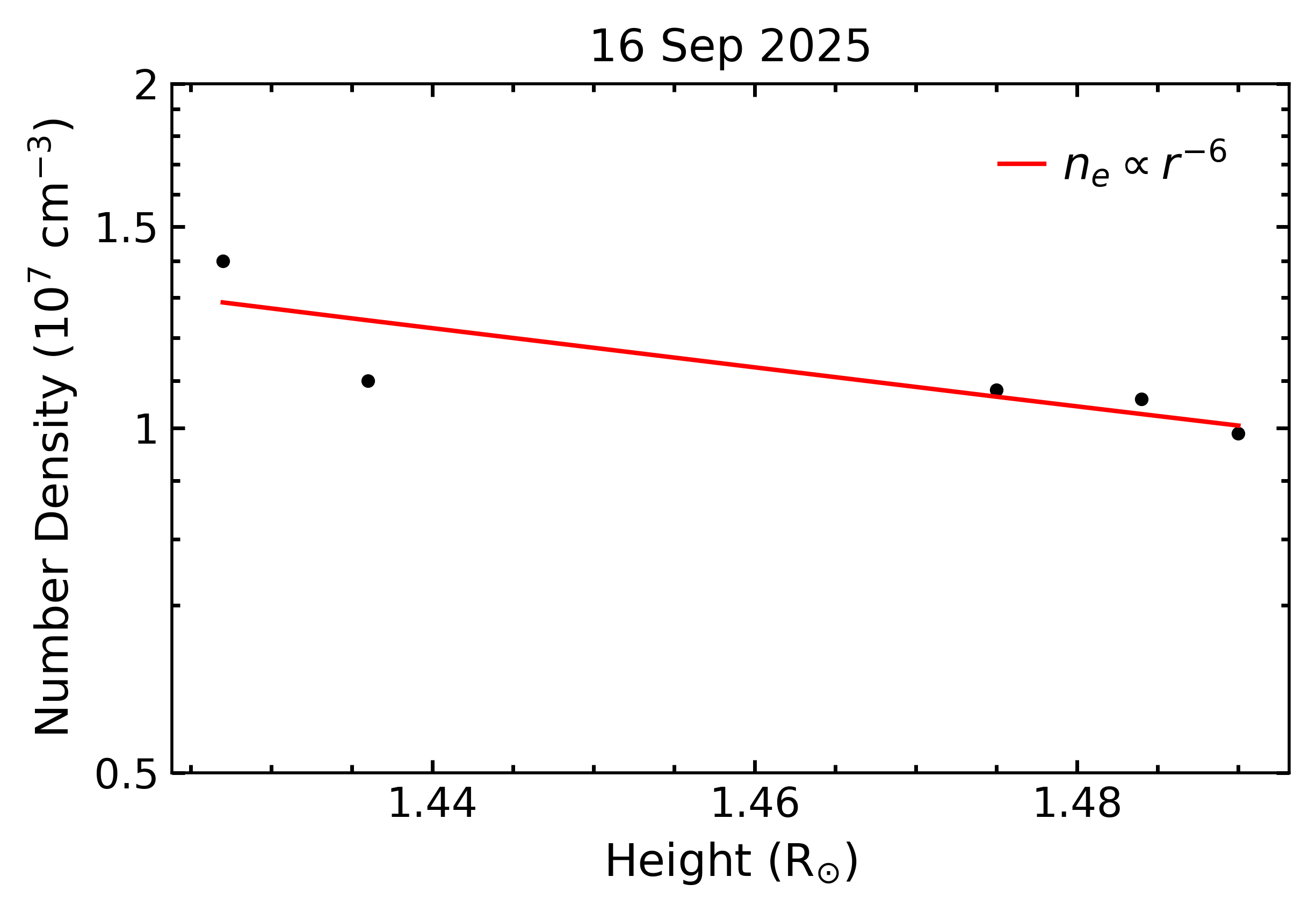}
\caption{Figure depicts the evolution of CME electron number density for the 14 Sep and 16 Sep CMEs in the left and right panels, respectively. Black circles denote the estimated number densities, and the red solid lines represent the fitted power-law profiles.}
\label{fig:numden_hei}
\end{figure*}

To directly compare the CME number density estimates with the VELC measurements, we back-extrapolate the observed number density of the 14 and 16 Sep CMEs from their first tracked heights to 1.09~$R_\odot$ and 1.12~$R_\odot$ (Section~\ref{sec:velc}), respectively, by using the fitted power-law relation and assuming it remains valid down to these heights. The power-law relations for both CMEs are obtained by fitting the observed electron number density as a function of heliocentric distance. Despite the relatively small number of data points, the fits, shown by the red solid lines in Figure~\ref{fig:numden_hei}, provide reasonably good representations of the observed density evolution, yielding $n_e \propto r^{-3.3}$ for the 14 Sep CME and $n_e \propto r^{-6}$ for the 16 Sep CME. The steeper density fall-off for the 16 Sep CME is consistent with its stronger lateral expansion at lower coronal heights compared to the estimates of the 14 Sep CME. Using the back-extrapolated number density and volume at 1.09~$R_\odot$ and 1.12~$R_\odot$ for the 14 and 16 Sep CMEs, respectively, we estimate the corresponding CME masses at these heights. The back-extrapolated parameters, along with their associated uncertainties, are listed in the ``Extrapolated'' row of Table~\ref{tab:compar_aspiics_velc}. We note that the back-extrapolated masses are slightly larger than the masses estimated at the first tracked heights, suggesting that the adopted power-law extrapolations may slightly overestimate the CME volume and/or number density at lower heights. Further studies using a larger sample of CMEs are required to investigate this in more detail. In the following section, these estimates are compared with the VELC measurements to examine how well the derived density profiles describe CME evolution in the low corona.

\subsection{CME Observations Using VELC and Comparison with ASPIICS} \label{sec:velc}

In this section, we estimate CME plasma parameters from VELC observations and compare them with the corresponding ASPIICS estimates described in Section~\ref{sec:aspiics}. VELC provides simultaneous spectroscopic observations using four straight slits oriented in the north–south direction (see Figure 2 of \citealt{Ramesh2024}), of which slit~1 and slit~4 observe the east and west corona, respectively, in either sit-and-stare or raster-scan mode. Both the 14 Sep and 16 Sep CMEs were observed on the east limb in sit-and-stare mode using slit~1. For both events, the center of the slit was located at a heliocentric distance of $\sim$1.06~$R_\odot$. The VELC observations are processed using procedures described in earlier studies \citep{Ramesh2024,Singh2025,Ramesh2025,Muthupriyal2025,Muthupriyal2025a}. The exposure time for both events is 4~s, with observational cadences of $\sim49$~s and $\sim53$~s for the 14 and 16 Sep events, respectively.

\begin{figure*}
\centering
\includegraphics[scale= 0.48,trim={0cm 0cm 0cm 0cm},clip]{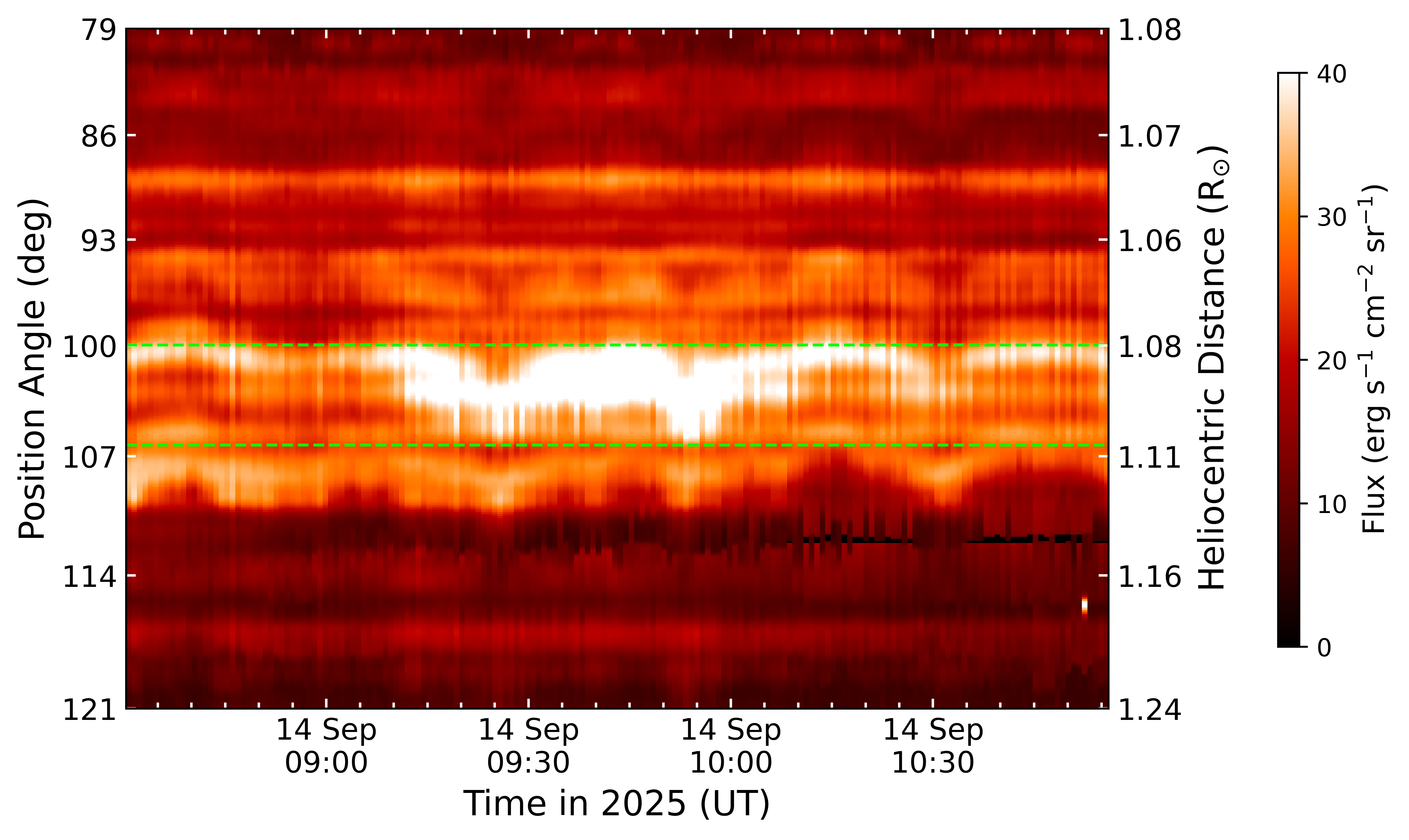}
\includegraphics[scale= 0.48,trim={0cm 0cm 0cm 0cm},clip]{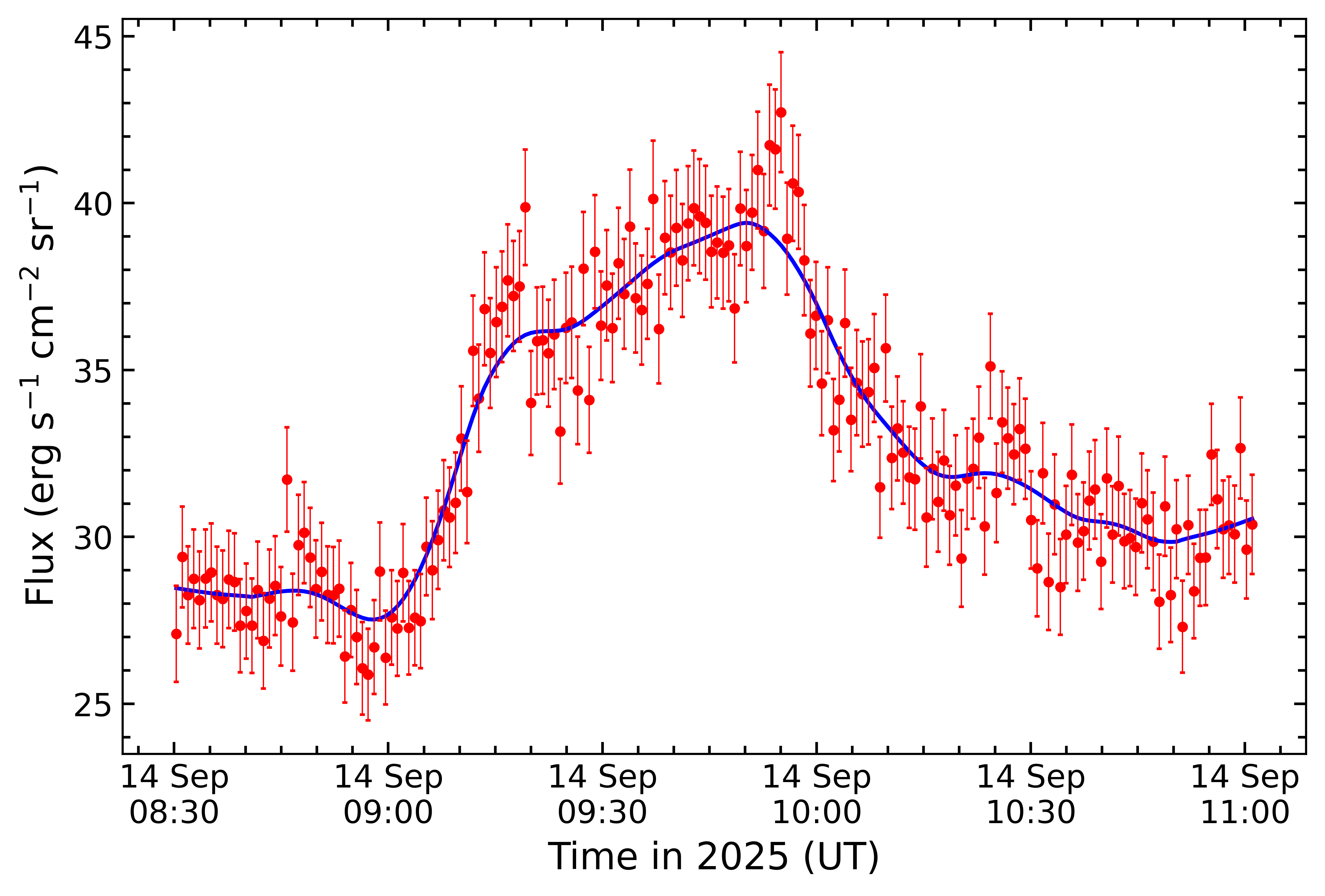}
\caption{Left: Temporal evolution of the coronal brightness in the 5303~$\AA$ emission line observed along slit~1 of VELC on 2025 Sep 14. The left y-axis shows the PA (measured from solar north), while the right y-axis shows the heliocentric distance of points on the slit corresponding to different PAs shown on the left y-axis. The enhanced-emission region between PAs of $\sim100^\circ$--106$^\circ$ is marked by dashed lime lines. Right: Corresponding flux averaged over the same PA range. The solid blue curve shows the flux smoothed using a local regression technique to suppress small-scale fluctuations while preserving the overall trend. The smoothed curve is shown only as a guide to the eye.}
\label{fig:14_sep_velc}
\end{figure*}

The left panel of Figure~\ref{fig:14_sep_velc} shows the total (integrated) Fe~XIV line intensity observed on 2025 September 14 (08:30--11:00~UT). The left y-axis shows the PA, measured from solar north, and the corresponding heliocentric distance of points on the slit is shown on the right y-axis. Enhanced emission is observed within the PA range of $\sim$100$^\circ$--106$^\circ$ (marked by the dashed lime lines in the southern hemisphere), corresponding to heliocentric distances of $\sim$1.08--1.11~$R_\odot$. The lateral extent of the emitting region can be estimated from the heliocentric distances of its starting PA of emission and ending PA of emission from the slit center, which is at 1.06~$R_\odot$ at the equatorial PA of 90$^\circ$. From the slit center, the lateral dimension of the starting PA will be $\sqrt{1.08^2-1.06^2}$ while the lateral dimension of the ending PA will be $\sqrt{1.11^2-1.06^2}$. The difference of these lateral dimensions corresponds to the lateral dimension of the enhanced emission region. The estimated lateral extent of the emission region is spread over $\sim$0.125 $R_\odot$. This emission can be represented (corresponds to the midpoint of the enhanced emission region) at the heliocentric distance of $\sim$1.09~$R_\odot$. This region spatially coincides with the CME observed in ASPIICS.

The right panel of the figure shows the flux averaged over the same PA range, shown by the red dots. The solid blue curve represents the flux smoothed using the Locally Weighted Scatterplot Smoothing (LOWESS) method, which performs locally weighted regressions to suppress small-scale fluctuations while preserving the overall trend. The smoothed curve is shown only as a guide to the eye; all quantitative analyses are performed using the observed data. The flux enhancement begins at $\sim08{:}56$~UT, consistent with the CME onset estimated from ASPIICS (Section~\ref{sec:cme_kinematics}), confirming that both instruments observed the same CME. The CME-related flux enhancement is obtained by subtracting the pre-event flux from the peak value, yielding $\sim$$12.5\pm2.2$~$\mathrm{erg~s^{-1}cm^{-2}sr^{-1}}$ (from $26.9 \pm 1.4$ to $39.4\pm1.7$~$\mathrm{erg~s^{-1}cm^{-2}sr^{-1}}$).

\begin{figure*}
\centering
\includegraphics[scale= 0.48,trim={0cm 0cm 0cm 0cm},clip]{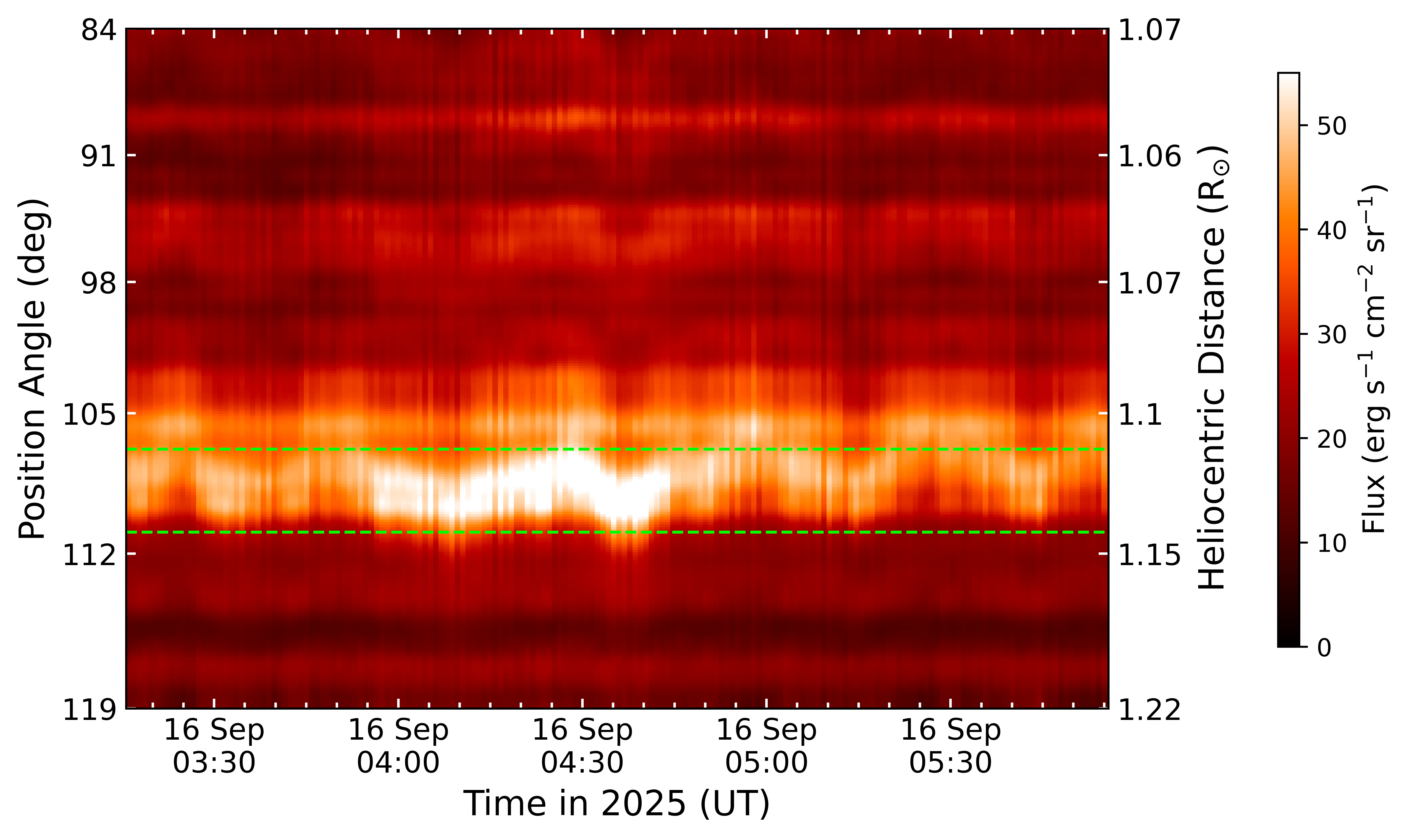}
\includegraphics[scale= 0.48,trim={0cm 0cm 0cm 0cm},clip]{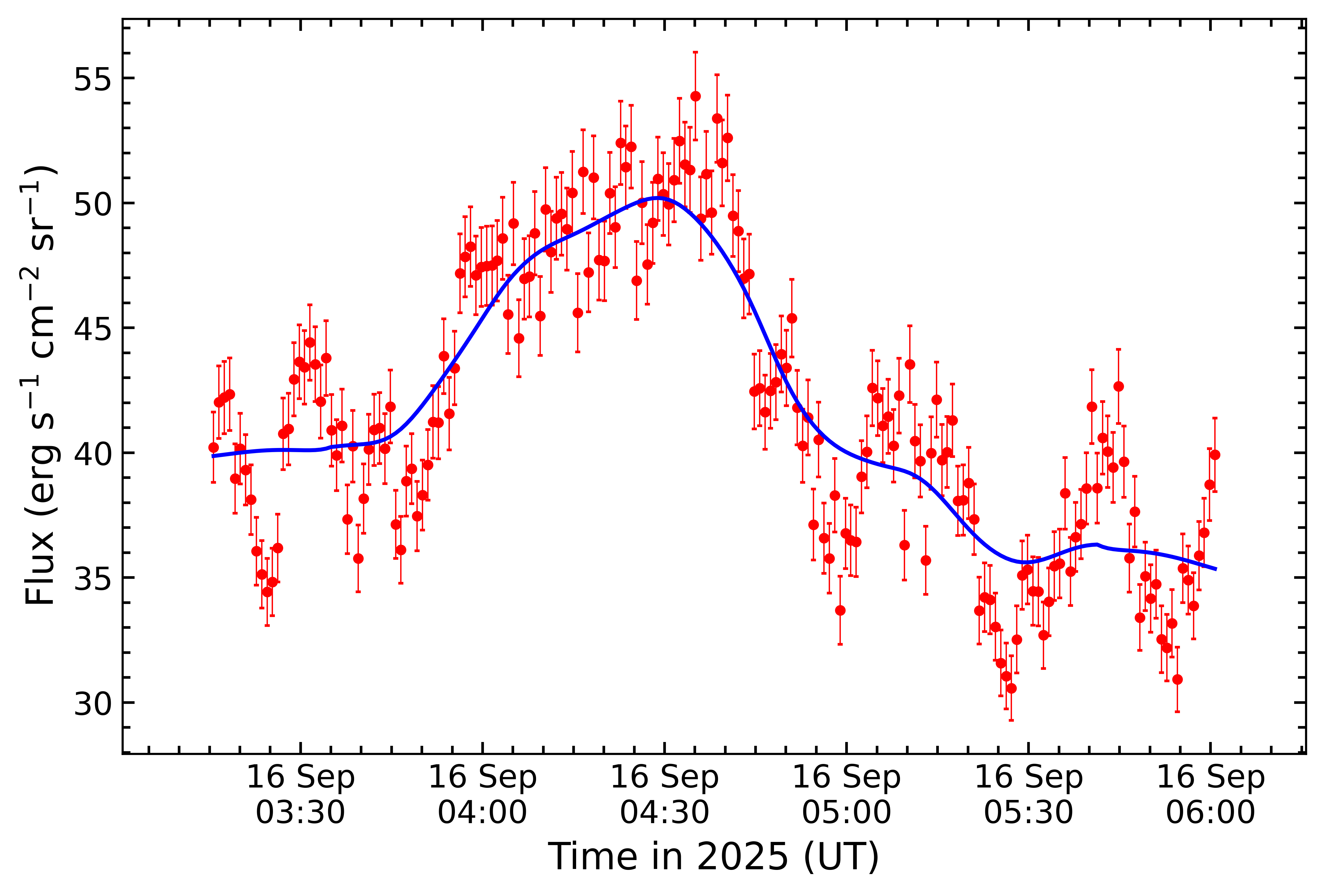}
\caption{Left: Temporal evolution of the coronal brightness in the 5303~$\AA$ emission line observed along slit~1 of VELC on 2025 Sep 16. The left y-axis shows the PA (measured from solar north), while the right y-axis shows the heliocentric distance of the points on the slit, corresponding to different PAs shown on the left y-axis. The enhanced-emission region between PAs of $\sim106^\circ$--112$^\circ$ is marked by dashed lime lines. Right: Corresponding flux averaged over the same PA range. The solid blue curve shows the flux smoothed using a local regression technique to suppress small-scale fluctuations while preserving the overall trend. The smoothed curve is shown only as a guide to the eye.}
\label{fig:16_sep_velc}
\end{figure*}

The left panel of Figure~\ref{fig:16_sep_velc} shows the total (integrated) intensity of the Fe~XIV line observed on 2025 September 16 (03:15--06:00~UT). The left y-axis shows the PA, measured from solar north, and the corresponding heliocentric distance of points on the slit is shown on the right y-axis. Enhanced emission is present over the PA range of $\sim106^\circ$--112$^\circ$. Similar to calculations done for 14 Sep CME, the emission region corresponds to a lateral extent of $\sim0.1~R_\odot$ at a representative heliocentric distance of 1.12~$R_\odot$. This is smaller than the lateral extent measured from the ASPIICS Fe~XIV imaging observations ($\sim0.32~R_\odot$ at $\sim1.38~R_\odot$). Assuming linear expansion with height, the corresponding extent at 1.12~$R_\odot$ would be $\sim0.25~R_\odot$, the order of magnitude consistent with the VELC estimate. Nevertheless, the remaining difference in the measured lateral extents suggests that the CME may have undergone over-expansion in the low corona.

The averaged flux (right panel) begins to increase at $\sim03{:}46$~UT, slightly earlier than the onset time estimated from ASPIICS, likely due to the assumption of constant speed in the ASPIICS-based extrapolation. We note that PAs over which the CME brightness is observed for VELC and ASPIICS are within $10^\circ$, and this indicates the detection of the same CME by both instruments. However, a slight offset in PA between both instruments could be possible due to non-radial propagation and overexpansion of the CME in the low corona. The corresponding flux enhancement is $\sim$$10.7\pm 2.2$~$\mathrm{erg~s^{-1}cm^{-2}sr^{-1}}$ (from $39.2\pm1.4$ to $49.9\pm1.7$~$\mathrm{erg~s^{-1}cm^{-2}sr^{-1}}$).

The electron density associated with the CME in the VELC observations is estimated by converting the total Fe~XIV (5303~$\AA$) line intensity ($I_\lambda$) of the CME plasma into emission measure (EM) using the CHIANTI contribution function ($G_{\lambda}$; \url{https://www.chiantidatabase.org/}), such that $\mathrm{EM}=I_\lambda/G_{\lambda}(T,n_e,r,\Phi_{disk})$ \citep{Landi2002,Young2003,Landi2009}, where $T$ is electron temperature, $n_e$ is electron density, and $\Phi_{disk}$ brightness of the solar disk governing incident radiation field. It usually assumes an isothermal plasma at the emission line's formation temperature \citep{Muthupriyal2025a}. The contribution function is a proportionality factor used to convert local plasma properties into a local emissivity at a given spectral line. Therefore, it contains the information about atomic physics, ionization balance, elemental abundance, and excitation mechanisms \citep{Boe2022}. In the solar corona, both collisional and radiative excitation contribute to the line emissivity, such that $\epsilon_{\lambda}=\epsilon_{\rm coll}+\epsilon_{\rm rad}$ and $G_{\lambda}=G_{\rm coll}+G_{\rm rad}$. The former scales as $n_e^2$, while the latter scales as $n_e$. In our study, we have used CHIANTI version 10 that also accounts for radiative excitation \citep{Boe2022}. In the solar corona, as heliocentric distance increases, density decreases faster than the solar disk radiation field; therefore, radiative excitation becomes increasingly important in the outer corona.

For the Fe~XIV 5303~$\AA$ line, whose contribution function peaks near a temperature of $1.8\times10^6$~K, the derived total contribution functions ($G_{\lambda}$), including both radiative and collisional excitation, are $2.0\times10^{-26}$~erg~s$^{-1}$~cm$^{3}$sr$^{-1}$ and $1.9\times10^{-26}$~erg~s$^{-1}$~cm$^{3}$sr$^{-1}$ for the 14 Sep and 16 Sep CMEs, respectively. The resulting EM values are $(6.2\pm1)\times10^{26}$~cm$^{-5}$ and $(5.5\pm1.1)\times10^{26}$~cm$^{-5}$ for the 14 and 16 Sep events, respectively. We note that the contribution function would be reduced by approximately $\sim$15\% if only collisional excitation is considered for the 14 Sep and 16 Sep CME observed at 1.09$R_\odot$ and 1.12 $R_\odot$, respectively.

The electron density is then derived using $n_e=(\mathrm{EM}/L)^{1/2}$, considering the line-of-sight depth ($L$) equals the estimates of CME lateral extent \citep{Muthupriyal2025a}, as described above. The estimated number densities and uncertainties are listed in the eighth column of Table~\ref{tab:compar_aspiics_velc}, where the uncertainties are derived through standard error propagation, also including a $5\%$ uncertainty in $L$. Assuming a conical geometry with the lateral extent equal to both the CME radius and height, the CME volume and corresponding uncertainties are listed in the seventh column. Using the VELC-derived enhanced emission and associated electron density, the CME mass is then estimated as $M=1.97\times10^{-24}n_eV$~g, assuming a plasma composition of 90\% hydrogen and 10\% helium, to make it consistent with the ASPIICS white-light mass estimates. The resulting masses are listed in the sixth column of Table~\ref{tab:compar_aspiics_velc}.

Comparison of the VELC-derived CME masses with the ASPIICS values, extrapolated to 1.09~$R_\odot$ for the 14 Sep CME and to 1.12~$R_\odot$ for the 16 Sep CME, shows good agreement. The extrapolated ASPIICS CME volumes and density differ from VELC estimates by a factor of $\sim3$ for both CMEs. The difference may arise from different geometries used for volume estimation in the ASPIICS and VELC observations, as well as from back-extrapolation of ASPIICS estimates. In the following section, we compare the thermal energy derived from VELC observations with corresponding estimates obtained from back-extrapolated ASPIICS parameters. We subsequently examine the evolution of the kinetic, thermal, and magnetic energies of the CMEs using ASPIICS observations, with the magnetic energy evolution estimated from observed CME properties under physically motivated assumptions.

\subsection{Estimating CME Energetics Using ASPIICS and VELC} \label{sec:comparing_energy}

Using the number density ($n_e$) and volume ($V$) derived from the VELC observations (Section~\ref{sec:velc}), we estimate the thermal energies of the 14 and 16 Sep CMEs at 1.09~$R_\odot$ and 1.12~$R_\odot$, respectively, using $E_{\rm th}=3n_ek_BTV$, where $k_B$ is the Boltzmann constant. For both events, a temperature of $1.8\times10^6$~K, close to the peak formation temperature of the Fe~XIV 5303~$\AA$ emission line, is assumed at the VELC observed height. This temperature is adopted in the absence of observations from other emission lines that could constrain the temperature distribution of the CME plasma. The estimated thermal energies and their uncertainties, derived through standard error propagation from $n_e$ and $V$, are listed in the third column of the top panel of the Table~\ref{tab:energy_aspiics_velc}. The difference in the VELC-derived thermal energies of the selected CMEs may indicate a genuine difference in their thermal energy content. Alternatively, it may partly arise from the assumption of a single, fixed peak-formation temperature of the Fe XIV line for both events, which does not account for the actual temperature distribution of the CME plasma.

We note that the VELC-derived thermal energies correspond to the Fe~XIV emitting component of the CME plasma and may therefore not fully represent the total thermal content of the CME. The thermal energy is also estimated using the extrapolated ASPIICS parameters $n_e$ and $V$ (see Section~\ref{sec:cme_parameters}) and is listed in the second column of the table. For both CMEs, despite differences in volume and density estimates from VELC and ASPIICS, there is good agreement in the thermal energy estimates from both observations. This suggests that the inferred CME density and volume from ASPIICS compensate each other and provide a reasonable estimate of CME thermal energy.

We make an attempt to infer the change in magnetic field before and during the eruption of both CMEs using the VELC observations. In an earlier study based on the green line observations, \citet{Wang1997} showed that $B \propto n_e^{1.25}$ considering intensity $I \propto n_e^2$. This would imply that $B \propto I^{5/8}$. Using the VELC-observed intensity enhancement for our events, we infer that the magnetic fields of the 14 Sep and 16 Sep CMEs during their eruptions will be $\sim$30\% and $\sim$20\% higher than their pre-event background, respectively.

\begin{table*}
\centering
\footnotesize
\renewcommand{\arraystretch}{1.5}
\begin{tabular}{cccc}
\hline
\multicolumn{4}{c}{Comparison of Thermal Energy (erg) from ASPIICS and VELC}\\
\hline
{Date} & Height ($R_\odot$) & {ASPIICS ($\sim$Error)} & {VELC ($\sim$Error)}\\
\hline
{14 Sep} & 1.09 & $1.1\times 10^{29}~(\pm 0.33)$ & $1.4\times 10^{29}~(\pm 0.25)$\\
{16 Sep} &1.12 & $4.1\times 10^{28}~(\pm 1.2)$ & $6.1\times 10^{28}~(\pm 1.1)$\\
\hline
\multicolumn{4}{c}{Energetics (erg) from ASPIICS}\\
\hline
\multirow{2}{2em}{Date} & {Range of $\mathrm{Kinetic~Energy~(\frac{1}{2}mv^2)}$} & {Range of $\mathrm{Thermal~Energy~(3{N_e}k_B T)}$} & {Range of $\mathrm{Magnetic~Energy~(\int \frac{B^2}{8 \pi}dV)}$}\\

 & {($\sim$Error)} & {($\sim$Error)} & {($\sim$Error)}\\
\hline
{14 Sep} & $4.96 \times 10^{29}~(\pm0.75) - 6.75\times10^{29}~(\pm1.0)$ &  $7.5 \times 10^{27}~(\pm 2.3) - 8.2\times10^{27}~(\pm2.5)$ &  $5.6 \times 10^{29}~(\pm2)- 8.7\times10^{29}~(\pm3.1)$ \\

{16 Sep} & $1.08 \times 10^{28}~(\pm0.16) - 1.13\times10^{28}~(\pm0.17)$ &  $6.18 \times 10^{27}~(\pm1.9) - 1.3\times10^{28}~(\pm0.4)$ &  $1.3 \times 10^{28}~(\pm0.48) - 3.6\times10^{28}~(\pm1.3)$ \\
\hline
\end{tabular}
\caption{The top panel of the table presents the thermal energy estimates of both CMEs from ASPIICS and VELC. The bottom panel lists the ranges (from the initial to the last tracked height) of kinetic, thermal, and magnetic energies. The associated uncertainties in the estimated parameters are given in square brackets.}
\label{tab:energy_aspiics_velc}
\end{table*}

\begin{figure*}
\centering
\includegraphics[scale= 0.76,trim={0cm 0cm 0cm 0cm},clip]{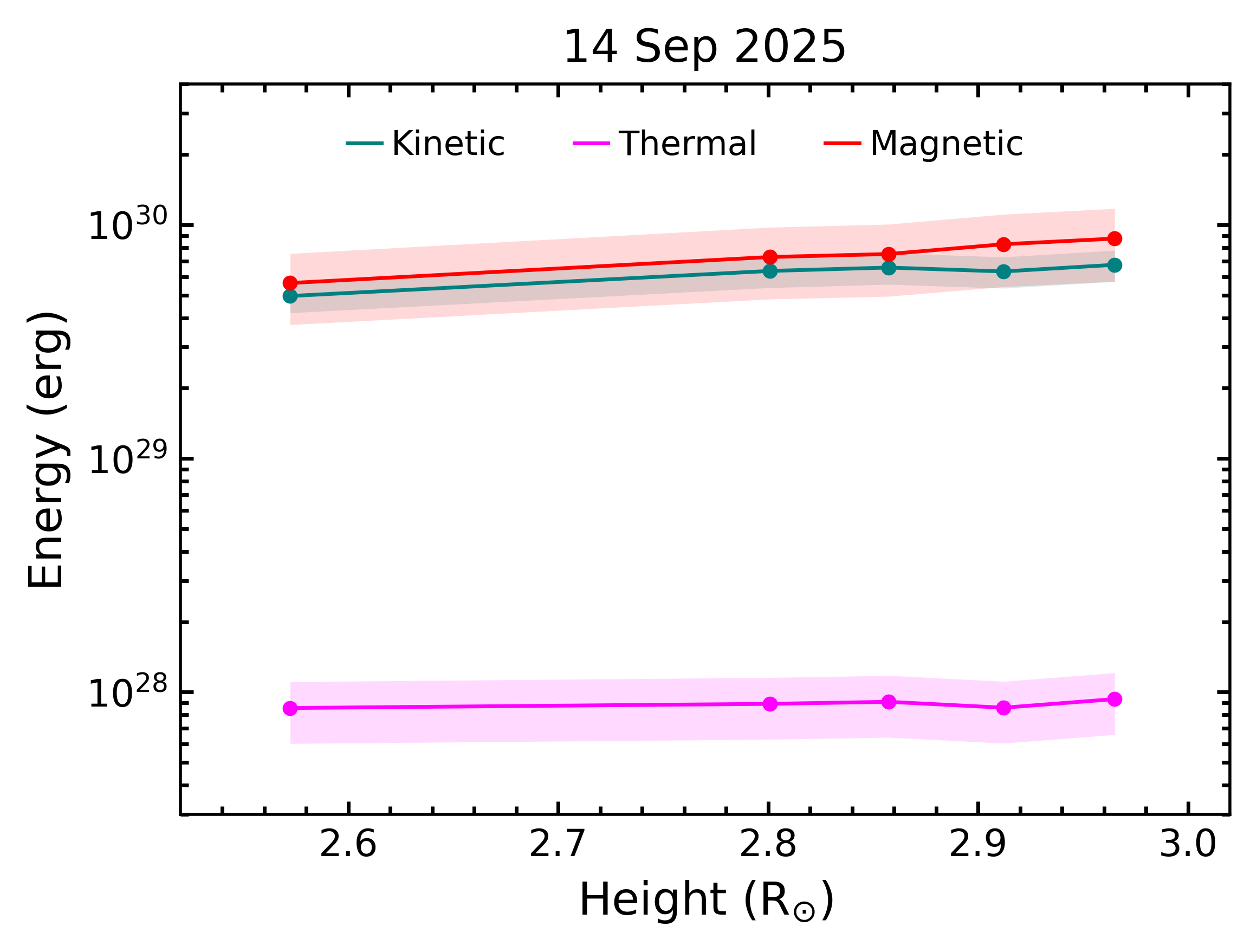}
\includegraphics[scale= 0.76,trim={0cm 0cm 0cm 0cm},clip]{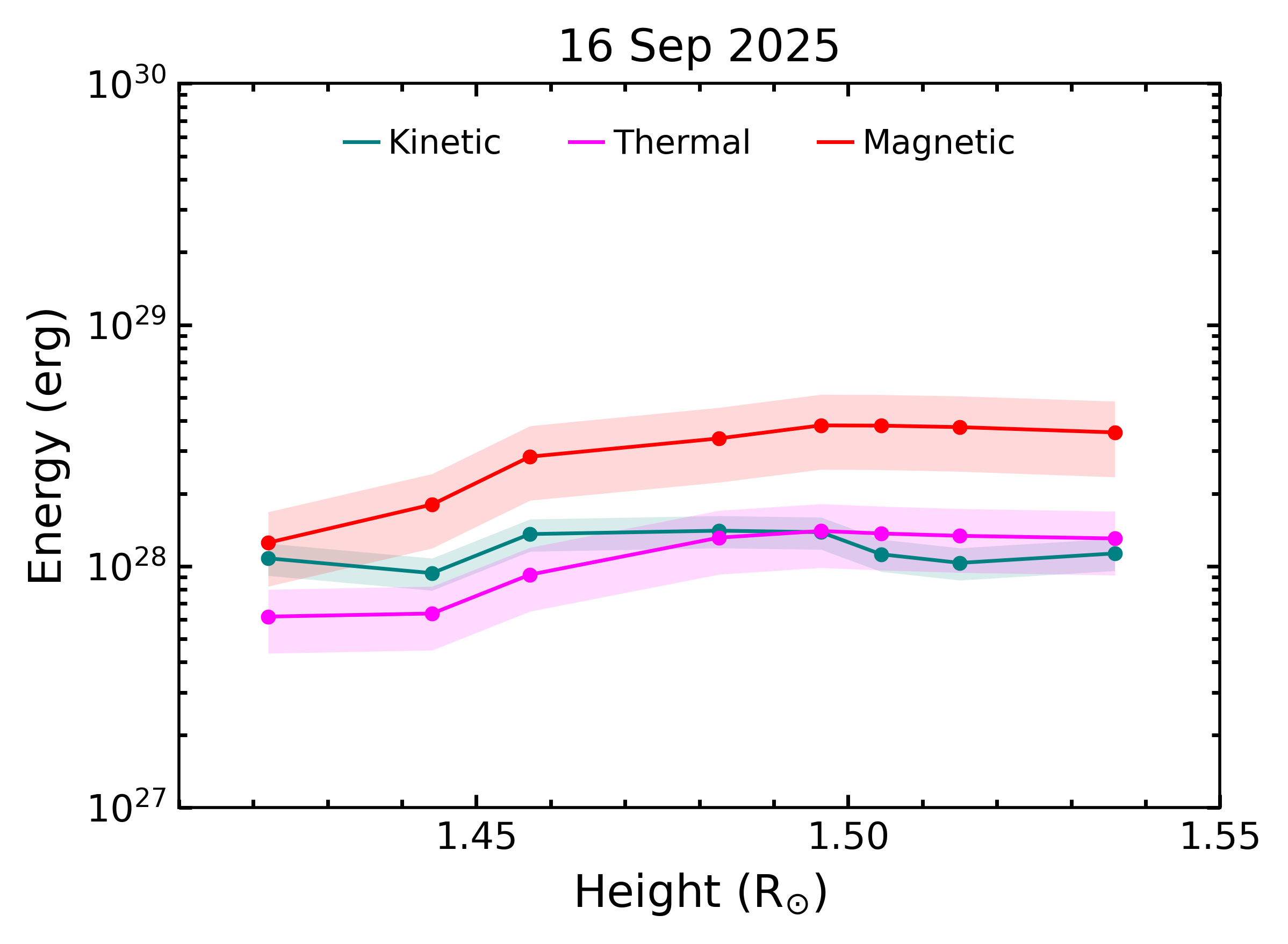}
\caption{Figure depicts the evolution of kinetic energy (teal), thermal energy (magenta), and magnetic energy (red) for the 14 Sep and 16 Sep CMEs in the left and right panels, respectively. The shaded areas with the same color represent the uncertainty in the estimates.}
\label{fig:energy}
\end{figure*}

We attempt to estimate the evolution of kinetic, thermal, and magnetic energies over the observed (from ASPIICS) height range of 2.57-2.97 $R_\odot$ and 1.43-1.54 $R_\odot$ for 14 Sep and 16 Sep CMEs, respectively. The kinetic energy is calculated from the CME mass and speed derived in Section~\ref{sec:cme_parameters}, using $E_{kin} = \frac{1}{2}mv^2$, with uncertainties estimated through standard error propagation. The resulting energy ranges, from the first to the last tracked heights, are listed in the bottom panel of Table~\ref{tab:energy_aspiics_velc}. Figure~\ref{fig:energy} depicts the evolution of kinetic energy in teal in the left and right panels for the 14 Sep and 16 Sep CMEs, respectively. For the 14 Sep CME, the kinetic energy gradually increases from $(4.96\pm0.75)\times10^{29}$ to $(6.75\pm1)\times10^{29}$~erg between the first and last tracked heights, despite a slight decrease in the CME speed. While the kinetic energy of the 16 Sep CME remains nearly constant, varying only from $(1.08\pm0.16)\times10^{28}$ to $(1.13\pm0.17)\times10^{28}$~erg, even though its speed decreases more rapidly over the observed height range. These trends indicate that for both events, the increase in CME mass offsets the decrease in CME speed, underscoring the importance of accurately estimating mass and speed, as the former may have larger uncertainties due to various factors as outlined in Section~\ref{sec:discussion}.

To infer the thermal energy evolution from the ASPIICS observations, the CME temperature is evolved from the VELC height to the ASPIICS observed heights using the polytropic relation $T\propto n_e^{\Gamma-1}$, where $\Gamma$ is the polytropic index. The temperature evolution starts from $1.8\times10^6$~K, corresponding to the peak formation temperature of the Fe~XIV line, adopted at the VELC height. Since previous studies suggest that CMEs near the Sun are in a heat-releasing state ($\Gamma \ge 1.66$) \citep{Khuntia2023,Khuntia2024}. Therefore, in our study, we assume a constant value of $\Gamma=1.8$ for both events for the observed height range of ASPIICS. The adopted value of $\Gamma$ yields lower temperatures than those expected under adiabatic evolution. We note that using an identical $\Gamma$ for both CMEs does not capture any true event-to-event differences in polytropic behavior, which may contribute to the derived differences in thermal energy evolution between the two CMEs. The corresponding thermal energies and uncertainties, derived through error propagation from the uncertainties in $n_e$ and $V$, are also listed in Table~\ref{tab:energy_aspiics_velc}.

Figure~\ref{fig:energy} depicts the evolution of thermal energy in magenta in the left and right panels for the 14 Sep and 16 Sep CMEs, respectively. For the 14 Sep CME, the thermal energy remains almost constant from $(7.5\pm2.3)\times10^{27}$ to $(8.2\pm2.5)\times10^{27}$~erg between the first and last tracked heights. While the thermal energy of the 16 Sep CME increases from its first to the last observed tracked height, varying from $(6.18\pm1.9)\times10^{27}$ to $(1.3\pm0.4)\times10^{28}$~erg. This reflects a balance between expansion-driven cooling and increasing mass for the 14 Sep CME, while for the 16 Sep CME, the increase in thermal energy may be primarily owing to its more rapid mass growth during expansion.

Since, in our study, direct measurements of the magnetic field within the CME are not available. Therefore, the magnetic energy evolution cannot be uniquely determined and must be inferred from physically motivated assumptions combined with observed CME properties. In the present work, we use the measured CME mass, volume, and density evolution to estimate the magnetic field evolution and, consequently, the magnetic energy. Using electron density estimates from the ASPIICS observations, we found that the density evolution follows a power-law dependence on heliocentric distance.

To find an approach for the magnetic-field evolution, we assume that the CME LE propagates at speeds below the local Alfv\'en speed ($V_A$) throughout the ASPIICS field of view. This is consistent with the absence of discernible CME-driven shock signatures and type II radio bursts \citep{Klein1982,Mann2003,Burgess2015}. This implies that in the inner solar corona, we can express $V_{CME}\leq V_A$ (sub-Alfv\'enic); where $V_{CME}$ is the CME speed relative to the upstream solar wind speed. If the Alfv\'en speed of the coronal background is assumed to be constant over the small observed height range (2.57-2.97 $R_\odot$ for 14 Sep CME and 1.43-1.54 $R_\odot$ for 16 Sep CME), then $V_A = \frac{B}{\sqrt{4\pi\rho}} \approx \mathrm{constant}$, which implies that $B \propto \rho^{1/2}$, where $\rho$ is the mass density. We obtain the minimum magnetic field consistent with the sub-Alfv\'enic condition by taking $V_A = V_{CME}$. This yields magnetic-field strengths of approximately 0.8~G (at 2.57 $R_\odot$) and 0.4~G (at 1.43 $R_\odot$) for the 14 Sep and 16 Sep CMEs, respectively. These values therefore represent the minimum magnetic field strengths consistent with the assumed sub-Alfv\'enic, shock-free propagation. If the actual Alfvén speed is higher than the CME speed, the corresponding magnetic field and magnetic energy would also be higher.

The estimated values of the magnetic field at the first observed are evolved at other tracked heights using the power law profile of the derived electron number density (Section~\ref{sec:cme_parameters}) through the relation $B\propto n_e^{1/2}$. If $n_e \propto r^{-m}$ then the magnetic field evolves as $B\propto r^{-0.5m}$. This yields magnetic field power-law profiles of $B\propto r^{-1.65}$ for the 14 Sep CME and $B\propto r^{-3}$ for the 16 Sep CME. The corresponding lower limits of the magnetic field (estimated using extrapolation) at $2~R_\odot$ are $\sim1.2$~G and $\sim0.15$~G for the 14 Sep and 16 Sep CMEs, respectively, in good agreement with values reported in earlier studies \citep{Maia2007,Sasi2014,Ramesh2021}.

Using the estimated magnetic-field values and CME volume, we made an attempt to estimate the lower limit of the CME magnetic energy ($E_{mag} = \int \frac{B^2}{8 \pi}dV$) at each tracked height. The associated uncertainties are obtained by propagating the volume uncertainty and by assuming uncertainties of 15\% in both the magnetic-field strength and its power-law exponent. The resulting magnetic energies and their uncertainties at the first and the last tracked height are listed in the bottom panel of the Table~\ref{tab:energy_aspiics_velc} and are shown in red in Figure~\ref{fig:energy}.

For the 14 Sep CME, the magnetic energy increases from $(5.6\pm2)\times10^{29}$ to $(8.7\pm3.1)\times10^{29}$~erg. While the magnetic energy of the 16 Sep CME increases from $(1.3\pm0.48)\times10^{28}$ and then reaches a constant value of $(3.6\pm1.3)\times10^{28}$~erg. The uncertainty in the magnetic energy reaches up to $\sim35\%$ for both CMEs, primarily due to the poorly constrained magnetic field strength, consistent with earlier studies \citep{Vourlidas2000}. Such uncertainties are expected to persist until direct measurements of CME magnetic fields become available in the inner corona.

For the 14 Sep CME, the magnetic and kinetic energies remain comparable throughout the observed height range, contributing $\sim$50--55 ($\pm$2)\% and $\sim$45 ($\pm$2)\% of the total energy, respectively, while the thermal energy accounts for only about 1\%. In contrast, the magnetic energy remains the dominant component for the 16 Sep CME, contributing $\sim$40--60 ($\pm$2)\% of the total energy, whereas the thermal energy contributes $\sim$20 ($\pm$2)\% and becomes comparable to the kinetic energy at larger heights. Compared to the 14 Sep CME, the kinetic energy of the 16 Sep CME is nearly two orders of magnitude smaller, primarily because of its lower propagation speed. For both CMEs, even the lower limit of the magnetic energy exceeds the thermal energy and is comparable to or larger than the kinetic energy, highlighting the magnetically dominated nature of CMEs during their early evolution. The comparatively larger thermal-energy contribution in the 16 Sep CME indicates a distinct energy partitioning between the two events, which may reflect a genuine difference in their thermodynamic evolution. However, since the energy estimates rely on simplifying assumptions, this difference may arise from the adopted estimation methods.

We note that, owing to the observational constraints, the estimated thermal and magnetic energies are all derived from common observed quantities, including the CME number density, volume, and leading-edge speed. Although these energies are therefore not entirely independent, they exhibit different functional dependencies on the observed parameters. Consequently, their comparison provides an approximation of the relative evolution of the different energy components and is as accurate as the assumption on the physical parameters of CMEs.

We emphasize that the magnetic energy estimates presented in our study should be regarded as approximate estimates rather than unique determinations. They are intended to constrain a physically plausible range of magnetic energies based on the observed CME density and expansion. Accordingly, the derived magnetic energies are interpreted primarily in terms of their relative evolution and contribution to the overall CME energy budget within the inner-coronal height range sampled by ASPIICS. Our analysis highlights the importance of investigating the CME energy partition in the inner corona.

\section{Discussion} \label{sec:discussion}

\subsection{CME Kinematics, Mass, and Density Evolution}\label{sec:cmkimade}

In this study, we present the first coordinated observations of two CMEs using ASPIICS aboard Proba-3 and VELC aboard Aditya-L1, providing a unique opportunity to investigate CME mass, number density, kinematics, and energetics in the low corona. The complementary white-light and spectroscopic observations enable independent estimates of key CME properties and provide a valuable framework for examining the early evolution and energy partitioning of CMEs in the inner corona.

The two CMEs exhibit markedly different kinematic evolution. The 14 Sep CME is a relatively fast event, propagating with an average LE speed of $\sim645$~km~s$^{-1}$ and showing only weak deceleration over the observed height range. In contrast, the 16 Sep CME is much slower, with an average LE speed of $\sim145$~km~s$^{-1}$, and exhibits a more pronounced decrease in speed. For both events, the CME onset times inferred from the ASPIICS observations agree within $\sim20$~min with the onset of enhanced Fe~XIV emission observed by VELC, confirming that both instruments tracked the same eruptive structures.

The 16 Sep CME exhibits a lower propagation speed in the Fe~XIV channel ($\sim90$~km~s$^{-1}$) than in the ASPIICS wideband observations ($\sim145$~km~s$^{-1}$). This difference is likely a consequence of the distinct plasma diagnostics: white-light observations trace the bulk electron density distribution of the CME, whereas Fe~XIV emission samples plasma within a restricted range of temperature and ionization conditions. Consequently, the two diagnostics may emphasize different plasma components and need not exhibit identical kinematics.

The estimated mass of both CMEs shows an increasing trend with height, consistent with earlier studies of CME mass evolution in the low corona \citep{Bein2013}. This increase is more pronounced for the 16 Sep CME, whose mass approximately doubles over the observed height range, likely because the event is tracked from a lower coronal height than the 14 Sep CME. Despite the decrease in CME speed during propagation, the kinetic energy of both events shows a slight increase, indicating that the increase in CME mass largely compensates for the reduction in speed. This highlights the importance of accurately estimating both the CME mass and speed when evaluating its kinetic energy. In this study, we adopt a representative mass uncertainty of 15\%, estimated from repeated contour selection. However, this value likely represents only a lower limit to the total uncertainty, as it does not account for additional sources of error, such as line-of-sight effects, background subtraction, assumptions about CME composition, and uncertainties in the radiometric calibration of the coronagraph \citep{Vourlidas2000,Colaninno2009}.

The column electron densities estimated for both CMEs at their first observed heights are approximately one order of magnitude higher than those reported by \citet{Vourlidas2010}. This difference is likely because our estimates correspond to CMEs observed in the low corona, whereas \citet{Vourlidas2010} derived average column electron densities for CMEs at heliocentric distances greater than $10~R_\odot$ using observations spanning an entire solar cycle. We find that the electron number density of the 16 Sep CME follows a steeper power-law evolution ($n_e \propto r^{-6}$) than that of the 14 Sep CME ($n_e \propto r^{-3.3}$). The steeper density fall-off inferred for the 16 Sep CME is also consistent with its comparable lateral extent at lower coronal heights, indicating stronger expansion during its early evolution. Such rapid expansion may lead to a faster decrease in plasma density during CME propagation.

\subsection{Implications of the Assumed CME Geometry}

The back-extrapolated ASPIICS volumes and electron number densities for both CMEs differ from the corresponding VELC estimates by a factor of $\sim$3. These differences likely arise from the different geometrical assumptions used to estimate the CME volume and from the back-extrapolation of the ASPIICS parameters. Additionally, different geometric assumptions lead to different volume estimates and, consequently, different absolute values of the derived physical parameters, although the relative ordering of the parameters remains unchanged. Therefore, the most appropriate geometric representation should be selected based on the observational constraints of each instrument. For the white-light ASPIICS observations, the projected area of the complete CME structure is available, allowing the volume to be estimated using a spherical-sector approximation. Adopting a conical geometry instead can increase/decrease the volume of selected CMEs by up to $\sim50\%$. Moreover, the conical approximation assumes equal radial and lateral extents of the CME, an assumption that is generally not valid for real CMEs, whereas the spherical-sector geometry does not require this assumption. In contrast, the VELC sit-and-stare observations sample only a narrow slit across the CME and do not provide a complete image of the CME or its angular extent, which is required to apply the spherical-sector geometry. Consequently, a simplified conical approximation is adopted for the VELC observations. Assuming the CME to be spherical would represent an even more simplified approximation and is generally not representative of the morphology of real CMEs. Such an assumption would increase the estimated CME volume by up to $\sim300\%$ compared to the conical approximation. Thus, although the absolute values of the CME volume depend on the adopted geometry, the chosen geometrical representations are the most appropriate for the respective observational constraints of ASPIICS and VELC.

\subsection{CME Energetics: Estimates, Assumptions, and Partitioning}

Despite the disagreement between the ASPIICS back-extrapolated volumes and electron number densities with VELC, the mass and thermal energy estimates derived from the back-extrapolated ASPIICS parameters agree well with the VELC estimates, as the differences in volume and density largely cancel in the calculations. In this context, the mismatch between the extrapolated number density of both CMEs and the corresponding VELC estimate could also suggest that an even steeper density fall-off than the derived power-law profiles may be required in the low corona. This possibility can be examined further using a larger sample of CMEs observed with ASPIICS. The mass, number density, and thermal energy estimates derived from VELC observations for both CMEs are consistent with earlier studies \citep{Muthupriyal2025a}.

It is important to note that the thermal energies derived from VELC observations correspond to the Fe~XIV-emitting component of the CME plasma. Consequently, if a significant fraction of the CME mass resides at temperatures outside the temperature range to which Fe~XIV is most sensitive, the inferred thermal energies may not fully represent the total thermal content of the CME. In the absence of a direct estimate of the mass-weighted CME temperature, this effect remains difficult to quantify. Furthermore, VELC observed intensity indicates that the magnetic field may increase by 30\% and 20\% during the 14 Sep and 16 Sep CME eruptions, respectively, relative to their pre-event background.

The evolution of the CME magnetic and thermal energies over the ASPIICS height range depends on the adopted assumptions. The largest source of uncertainty in the CME energy budget arises from the magnetic energy estimates, which rely on the assumed evolution of the CME magnetic field. In this work, the magnetic-field evolution is derived assuming  $B\propto n_e^{1/2}$. To assess whether this approach is reasonable, we obtain an independent estimate based on the assumptions of frozen-in magnetic flux and self-similar CME expansion. This alternative approach assumes that the magnetic flux enclosed within the CME is conserved during its evolution. Consequently, it is valid only in the absence of significant magnetic reconnection or other processes that add or remove magnetic flux from the CME. Conservation of magnetic flux, $\Phi = BA = \mathrm{constant}$, combined with self-similar expansion and conservation of mass, yields $B \propto \rho^{2/3}$. For a density profile $n_e \propto r^{-m}$, this corresponds to $B\propto r^{-0.67m}$, whereas the adopted approach gives $B\propto r^{-0.5m}$. Thus, for the same density profile, the magnetic-field exponent is constrained to lie between m/2 and 2m/3. Thus, although these two approaches are based on different physical considerations, they yield nearly similar magnetic-field scalings over the observed height range. It suggests that we can rely on the evolution of the magnetic field to obtain a reasonable estimate of the lower limit of magnetic energy at the observed coronal heights of ASPIICS.

The thermal energy estimates depend on the assumed polytropic index because our understanding of the physical processes governing the heat budget is limited. Using an identical $\Gamma$ for both CMEs does not account for possible event-to-event differences in their thermodynamic evolution and may contribute to the differences in the estimated thermal energies. Accounting for the uncertainties in these assumptions through maximum error propagation yields uncertainties of up to $\sim35\%$ for the magnetic energy and $\sim30\%$ for the thermal energy. Therefore, the inferred magnetic energy evolution should be interpreted with appropriate caution. Such uncertainties are currently unavoidable due to the lack of direct measurements of CME magnetic field strength and plasma temperature in the low corona.

Despite these limitations, the estimated lower limits of the magnetic energy provide valuable insight into CME energy partitioning during the early stages of propagation. The two events exhibit markedly different energy partitioning. For both CMEs, the lower-limit magnetic energy remains the dominant energy component throughout the observed height range. In contrast, the thermal energy of the 14 Sep CME contributes only a negligible fraction of the total energy, whereas for the 16 Sep CME it increases during propagation and slightly exceeds the kinetic energy at larger heliocentric distances. This comparatively larger thermal-energy contribution in the 16 Sep CME may reflect a genuine difference in the thermodynamic evolution of the two events; however, since the energy estimates rely on simplifying assumptions, it may instead partly arise from the adopted estimation methods. Nevertheless, these differences highlight the diversity of CME thermodynamic and magnetic evolution in the low corona. For both CMEs, the thermal energy estimates show a reasonable match with the study of \citet{Bemporad2022}.

For the 14 Sep CME, at the first tracked height of 2.57~$R_\odot$, the thermal energy is only $\sim$1.5\% of the kinetic energy and $\sim$1.3\% of the magnetic energy, which agrees with the earlier studies \citep{Emslie2012,Gopalswamy2015a}, while the kinetic energy is comparable to the magnetic energy. In contrast, for the 16 Sep CME, at its first tracked height of 1.43~$R_\odot$, the thermal energy is $\sim$55 $(\pm2)$\% of the kinetic energy, consistent with earlier studies of CME energetics \citep{Hannah2013,QZhang2023}, and 50 $(\pm2)\%$ of the magnetic energy, while the kinetic energy is $\sim$85 $(\pm2)$\% of the magnetic energy. These results suggest that the two CMEs exhibit distinctly different energy partitioning in the low corona. 

The evolution of the energies is subject to the observational constraints and assumptions adopted for the CME physical parameters. Although the estimates of energies depend on common observed quantities, including the CME number density, volume, and LE speed, they have different functional dependencies on these parameters. Their comparison, therefore, provides useful insight into the relative evolution of the different energy components, while the reliability of the derived values depends on the validity of the adopted physical assumptions.

\subsection{Additional Observations and Study Limitations}\label{sec:limitstudy}

Although the primary focus of the 16 Sep CME was its energetics, we note a fast, wide-propagating structure ahead of it, marked by the purple curve in the rightmost panel of the middle row in Figure~\ref{fig:tracked_height_16_sep}, which may require a separate investigation. While its morphology may resemble a shock-like structure, the absence of contemporaneous radio signatures (\url{https://www.solarmonitor.org/}) makes such an interpretation uncertain. We think that the structure could instead correspond to another CME projected along a similar line of sight, possibly from a far-side source region.

Our study is subject to the limitations associated with the relatively weak nature of the selected CMEs and the simplifying assumptions adopted in the absence of direct measurements of key CME parameters, particularly the magnetic field and plasma temperature. Direct measurements of these quantities in future observations would provide stronger constraints on CME energetics in the inner corona. Nevertheless, to the best of our knowledge, this work provides the first coordinated determination of CME mass, number density, and energetics using simultaneous observations from Proba-3/ASPIICS and Aditya-L1/VELC, demonstrating the unique scientific potential of these complementary instruments for investigating CME energetics in the inner corona. Such coordinated observations provide direct constraints on CME mass, density evolution, and energy partition near the Sun, offering new insights into the early evolution of CMEs. Future coordinated observations, particularly when combined with near-Sun in situ measurements, may enable tighter constraints on the magnetic, thermodynamic, and density evolution of CMEs. Extending this analysis to a larger sample of coordinated ASPIICS and VELC observations will provide a more comprehensive understanding of CME energetics and the physical processes governing their early evolution.

\section{Conclusions} \label{sec:conclusion}

\begin{itemize}

\item This study presents the first coordinated observations of CMEs using Proba-3/ASPIICS and Aditya-L1/VELC, demonstrating the scientific potential of combining white-light coronagraphic and spectroscopic observations to constrain CME mass, density, and energetics in the inner corona, a region that has remained difficult to probe with earlier instrumentation.

\item Despite relying on different observational constraints and correspondingly different geometric assumptions, the ASPIICS and VELC derived mass and thermal energy estimates show good mutual agreement. This consistency lends confidence to the coordinated observation approach as a reliable, cross-validated route to constraining CME properties in the inner corona. However, this apparent agreement should be interpreted with caution, as it could arise because the back-extrapolated ASPIICS volumes and densities individually differ from the VELC estimates by a factor of $\sim$3, with the differences in both largely canceling in the mass and thermal energy calculations.

\item The two CMEs studied here exhibit markedly different energy partitioning: one event shows kinetic and magnetic energies of comparable magnitude with a negligible thermal contribution, while the other shows a substantial and growing thermal energy component that becomes comparable to its kinetic energy at larger heights. This diversity may indicate that CMEs need not follow a general trend of energy partitioning. However, the derived energetics necessarily rely on simplifying but physically motivated assumptions which are made in the absence of direct measurements of the CME magnetic field and plasma temperature at these heights. The observed differences in energy partitioning between the two events should therefore be interpreted with caution, as they may reflect genuine physical differences, the limitations of the adopted methods, or both.

\item Our study is limited by the relatively weak nature of the selected CMEs and by the simplifying assumptions required in the absence of direct measurements of key CME parameters. Despite these limitations, this study provides physically motivated, approximate estimates of CME energy partitioning in the inner corona, offering a useful first step toward characterizing CME evolution in this observationally challenging region. Addressing these limitations further will require extending this coordinated observational approach to a larger sample of CMEs, including stronger events, and combining it with near-Sun in-situ measurements; such efforts will be essential for building more general and robust constraints on CME energy partitioning in the inner corona, and for clarifying the physical processes that govern this partitioning as CMEs emerge into the heliosphere.

\end{itemize}

\begin{acknowledgements}
We acknowledge P. Savarimuthu, S. Nagashree, and E. Yuvashree for their team efforts at the VELC Payload Operations Center in processing VELC data. The ASPIICS data are courtesy of the Proba-3/ASPIICS consortium. Proba-3 is a technology demonstration mission of the European Space Agency (ESA) and a Mission of Opportunity in the ESA Science Programme. The ROB team thanks the Belgian Federal Science Policy Office (BELSPO) for the provision of financial support in the framework of the PRODEX Programme of ESA under contract numbers 4000145189 and 4000147286. We thank the team of SOHO/LASCO for making their observational data publicly available. P.L. acknowledges financial support from the Centre national d'études spatiales (CNES) in the framework of the Proba-3 mission. S.G. acknowledges the support from grant 25-18282S of the Czech Science Foundation (GA\v CR).
\end{acknowledgements}

\end{document}